\documentclass[11pt,a4paper]{article}
\usepackage[utf8]{inputenc}
\usepackage[english]{babel}
\usepackage{amsmath}
\usepackage{bbm}
\usepackage{amsfonts}
\usepackage{amssymb}
\usepackage{makeidx}
\usepackage{graphicx}
\usepackage{lmodern}
\usepackage{url}
\usepackage[title,titletoc,toc]{appendix}
\usepackage[pdfpagemode={UseOutlines},bookmarks=true,bookmarksopen=true,
   bookmarksopenlevel=0,bookmarksnumbered=true,hypertexnames=true,
   colorlinks,linkcolor={green!10!blue!80!black},citecolor={red!40!blue}, urlcolor=black!60!white,
   pdfstartview={FitV},unicode,breaklinks=true]{hyperref}
\usepackage[square,numbers,comma]{natbib}
\usepackage{tikz}
\usepackage{hyperref}

\usepackage[left=2cm,right=2cm,top=2cm,bottom=2cm]{geometry}
\author{Debajyoti Sarkar, Niladri Paul\footnote{Debajyoti Sarkar, Niladri Paul did a significant amount of work presented in this article when they were in the Department of Physics, Indian Institute of Technology, Kanpur 208016, India.} \thanks{Email:~ debajyoti, npaul @iucaa.in}\\
\normalsize Inter University Centre for Astronomy and Astrophysics, Pune 411007, India 
\\Kaushik Bhattacharya, Tarun Kanti Ghosh \thanks{kaushikb, tkghosh@iitk.ac.in} \\
\normalsize Department of Physics, Indian Institute of Technology, Kanpur 208016, India}
\title{An Effective Hamiltonian Approach to Quantum Random Walk}
\usepackage{duerer}
\usepackage[T1]{fontenc}
\date{}
\usepackage{blindtext}
\begin{document}
\maketitle
\begin{abstract}
In this article we present an effective Hamiltonian approach for
Discrete Time Quantum Random Walk. A form of the Hamiltonian for one
dimensional quantum walk has been prescribed, utilizing the fact that
Hamiltonians are the generators of time translations. Then an attempt
has been made to generalize the techniques to higher dimensions. We
find that the Hamiltonian can be written as the sum of a Weyl
Hamiltonian and a Dirac comb potential. The time evolution operator
obtained from this prescribed Hamiltonian is in complete agreement
with that of the standard approach. But in higher dimension we find
that the time evolution operator is additive, instead of being
multiplicative like that of Ref.~\citep{Chandrashekar_2013}. We showed that in
case of two-step walk, effectively the time evolution operator can
have multiplicative form.  In case of a square lattice, quantum walk
has been studied computationally for different coins and the results
for both the additive and the multiplicative approaches have been
compared. Using the Graphene Hamiltonian the walk has been studied on
a Graphene lattice and we conclude the preference of additive approach
over the multiplicative one.
\end{abstract}

\section{Introduction}\label{intro}
Quantum walk (QW) is a formulation implementing quantum principles on the
classical random walk problem. The classical version in one dimension can be described
with a two state coin. As the particle reaches a lattice point, the
coin is tossed and the particle moves towards either $+x$ or $-x$
direction based on the outcome of the toss. But unlike the classical,
if we include quantum concept, the particle would be in a
superposition of $+x$ and $-x$ state in general and the random walk
acquires a new nature as described in Ref.~\citep{Aharonov_1993},
\citep{Molinasmata_1996}, \citep{Nayak_Vishwanath_2000},
\citep{Ambainis_et_al_2001}, \citep{Kempe_2003} and others. Quantum
walk on a graph with different coins has been studied in Ref.~
\citep{Aharonov_2001},  \citep{Kempe_2003},
\citep{Kempe_2005}, \citep{Tregenna_2003} in great
detail. Quantum random walk has turned out to be very useful in
developing algorithms for quantum computation as studied in Ref.~\citep{Farhi_1998},
\citep{Wang_2001}, \citep{Shenvi_2003}, \citep{Ambainis_2003},
\citep{Childs_2003} and \citep{Childs_2009}.  \\ 
\indent In Ref.~\citep{Ambainis_et_al_2001},\citep{Nayak_Vishwanath_2000}, the problem of
quantum walk on a line was studied in Fourier space and an analytic
solution was obtained in the asymptotic limit. The effect of absorbing
boundaries was discussed in Ref.~\citep{Ambainis_et_al_2001},\citep{Meltem_2011}. In case of classical
random walk, the distribution attains a Gaussian nature asymptotically in time with a central peak and the variance is proportional to time;
whereas in case of quantum walk, the distribution spreads out from the
centre and the variance is proportional to the square of time as discussed in Ref.~\citep{Nayak_Vishwanath_2000}, \citep{Kempe_2003}. Quantum walk in two dimension has
been discussed in case of square and triangular lattice with various
coins like Hadamard, Grover, DFT etc. in Ref.~\citep{Stefanak_2008},
\citep{Stefanak_2010}, \citep{Chandrashekar_2013} and others. In this case also the distribution of the probability of finding the particle spreads away from the centre in general. \\ \indent In recent years,
experiments have been performed to implement discrete time quantum
random walk in case of NMR Quantum Information Processor cf. Ref.~\citep{Ryan_2005}, \citep{Du_2003}, trapped atoms cf. Ref.~\citep{Karski_2009}, ions cf. Ref.~\citep{Schmitz_2009}, \citep{Zahringer_2010}
and photons cf. Ref.~\citep{Schreiber_2010}, \citep{Peruzzo_2010}. The use of
Dirac-like Hamiltonian in the context of unitary quantum cellular
automata was discussed in Ref.~\citep{Birula_1994}. Recently the
Hamiltonian of the Discrete Time Quantum Random Walk (DTQRW) has been
studied in Ref.~\citep{Chandrashekar_2013}. \\

The material in the article is presented in the following way. In Sec.~\ref{concept:QW} we briefly introduce the concept of the quantum walk. In Sec.~\ref{sec: hamiltonian} we give a prescription to formulate the Hamiltonian of a QW system from a physical point of view and thereby obtain the time evolution operator from that Hamiltonian in Sec.~\ref{sec:Time evolution operator from a Dirac like Hamiltonian in 1 D}. We extend our Hamiltonian approach for two dimensional
lattice in Sec.~\ref{sec : qrw_square_lattice} and it is found that the evolution operator is additive in nature instead of being
multiplicative as mentioned in
Ref.~\citep{Chandrashekar_2013}. The other difference with
Ref.~\citep{Chandrashekar_2013} is the way in which the
effective Hamiltonian is perceived. In the above reference the
effective Hamiltonian does not have separate information about coin
operation and lattice translation, where as in our approach the
effective Hamiltonian shows a clear distinction between the two
separate operations. If we take the evolution operator to be
multiplicative, the walk becomes a two step walk; i.e. the particle must move along $Y$ direction after moving along $X$ direction and vice versa. Two consecutive moves along the same direction is forbidden.  But in our approach, it becomes a single step walk, i.e. the particle is allowed to move along the same direction in two consecutive steps. No path is forbidden. This notion of two-step and single step walk has been discussed in detail in Sec.~\ref{sec : qrw_square_lattice}. In  Sec.~\ref{sec: plot_square_lattice}, we compare the results of both approach
computationally in case of a square lattice using different quantum
coins. In the Sec.~\ref{sec:
  qrw_graphene}, the quantum walk on a Graphene lattice has been
studied starting from the Graphene Hamiltonian described in Ref.~\citep{Castro_Neto_2009}. In this case also we have presented a
comparative study of the additive and multiplicative approach.

\section{Concept of Quantum Walk} \label{concept:QW}

In case of quantum walk, the particle can be in a superposition of the available basis coin states
instead of a definite one as in the case of a classical random
walk. In one dimension this is equivalent to a two level problem. The two basis states of the coin can be represented by the kets $\vert +\rangle$
and $\vert -\rangle $. We assume that $\vert +\rangle$ is a
 coin state  for which the quantum particle moves in the $-x$
direction where as  for the other coin state $\vert -\rangle$
the particle moves in the $+x$ direction. A general coin state can be
represented as
 \begin{eqnarray} \label{eq1}
 \vert \chi \rangle &=& c_+\vert +\rangle + c_-\vert -\rangle .
 \end{eqnarray}
As soon as the particle reaches any lattice point, the state of the coin is changed; e.g. if the coin state of the particle was in an eigenstate ($\vert \pm \rangle  $), it changes to a superposed state of them. To include
superposition we need an operator which operates on $\vert +\rangle$
and $\vert -\rangle$ and results in a superposed state. Hadamard coin described in Ref.~\citep{Nayak_Vishwanath_2000} is a commonly used mixing operator in this regard. The effect of Hadamard operator on a coin state can be described
as
\begin{eqnarray} 
H_c \vert +\rangle &=& \frac{1}{\sqrt{2}}(\vert +\rangle +\vert -\rangle) ,
\label{eq2} \\
H_c \vert -\rangle &=& \frac{1}{\sqrt{2}}(\vert +\rangle -\vert -\rangle) .
\label{eq3}
\end{eqnarray}
In the following section, we discuss the Hadamard walk in brief in
position space formalism.
%%%%%%%%%%%%%%%%%%%%%%%%%%%%%%%%%%%%%%%%%%%%%%%%%%%%%%%%%%%%%
\subsection{Position Space Formulation} \label{method_2}
Following Ref.~\citep{Ambainis_et_al_2001} and \citep{Nayak_Vishwanath_2000}, we
consider the basis state of the particle as a product state of the chirality and the
position basis states : $\vert \pm \rangle \vert x\rangle$. Therefore we can
define a translation operator $T$ similar to that of Ref.~\citep{Chandrashekar_2013}, as
\begin{eqnarray} 
 T(n) &=& \vert + \rangle \langle + \vert \otimes \vert n-1 \rangle 
\langle n \vert +\vert - \rangle \langle - \vert \otimes \vert n+1
\rangle \langle n \vert ; \nonumber
\end{eqnarray}
where $n$ is the index of the lattice points. Here we have assumed that at each step the particle moves from one lattice point to the nearest lattice points only and the lattice spacing is unity.
Now in a basis of $\vert+\rangle \otimes \mathbbm{1}_N$ and
$\vert-\rangle \otimes \mathbbm{1}_N$, $T(n)$ has the form,
\begin{eqnarray}
T(n) &=&\begin{pmatrix}
\vert n-1 \rangle \langle n \vert &0\\0& \vert n+1 \rangle \langle n \vert 
 \end{pmatrix} . \label{eq7}
 \end{eqnarray}
 Here, $\mathbbm{1}_N$ \footnote{In our convention, $\mathbbm{1}_n$ (where $n \in N$, the set of natural numbers) denotes $n \times n$ identity operator in finite dimensional space, $\mathbbm{1}$ denotes infinite dimensional identity operator acting on the space spanned by the coordinates of all the points in space and $\mathcal{I}$ denotes infinite dimensional identity operator such that $\mathcal{I}=\mathbbm{1}_n \otimes \mathbbm{1}$} is the identity matrix in $N$ dimension whereas
 $N$ denotes the total number of discrete lattice points. Hence the
 time evolution operator is given by
 \begin{eqnarray} 
 W(n) &=& T(n)H_{c} \nonumber \\
 &=& \frac{1}{\sqrt{2}}
 \begin{pmatrix}
 \vert n-1 \rangle \langle n \vert &\vert n-1 \rangle \langle n
 \vert\\ 
\vert n+1 \rangle \langle n \vert &-\vert n+1 \rangle \langle n \vert
 \end{pmatrix}. \label{eq8} 
 \end{eqnarray}
Here $H_c$ is the Hadamard coin. In general, in place of $H_c$, we
can use any other quantum coin. The above one is the expression of
the time evolution operator if the particle was at the $n$-th lattice
point. Since the particle could be in any of the lattice points,
the total time evolution operator will be
\begin{eqnarray} 
\mathbb{W} = \sum_n W(n) . \label{eq9}
\end{eqnarray}
This is the time evolution operator for 1-D walk in case of position
space formulation.
%%%%%%%%%%%%%%%%%%%%%%%%%%%%%%%%%%%%%%%%%%%%%%%%%%%%%%%%%%%%%%%%%%%%%%%%%

\section{Hamiltonian Method} 
\label{sec: hamiltonian}
In this section we develop the Hamiltonian formulation of quantum
walk. In case of quantum walk the time evolution operator arises from two processes. First at the lattice points, the chiral state of the
particle changes and for that, at a given lattice point $n$, one can
write the chirality flipping operator as $W_1(n) = S \otimes \vert n
\rangle \langle n \vert$. Here $S$ is a $2 \times 2$ unitary operator. Then one uses a translational operator
$T(x)$ which moves the particle from one position to another position in
space. Clearly, $T(x)$ is a continuous operator acting at any point
$x$ in space (not necessarily lattice points). In case of the lattice points, the time evolution operator for a single step is given
by, $W(n) = T(n)W_1(n)$.  At places other than the lattice points $T(x)$ can itself act as the
time evolution operator since at those places, the chiral state of the particle remains unchanged. This perception of $W(n)$ immediately
leads to the concept of an effective Hamiltonian for the system
which acts as the generator for time evolution. Like any other
traditional Hamiltonian this effective Hamiltonian should be Hermitian.

Since the one dimensional quantum walk problem is a two level problem,
we expect the Hamiltonian to be proportional to a $2 \times 2$ matrix. We expect the wave function of the particle to be just translated in between two lattice points remaining unchanged in its form. Hence we consider the Hamiltonian proportional
to $kv$ where$p=\hbar k$ and $v$ is the effective velocity of the quantum
particle, not to be confused with the Lagrangian velocity. A straightforward calculation
shows that when the Hamiltonian $kv$ operates on some wave function,
it just translates the state keeping the form of wave function
invariant. Hence the Hamiltonian representing translation must be
proportional to $kv$. So we expect the Hamiltonian to be of the form
$\alpha kv \otimes \mathbbm{1}$ in between two
lattice points and $\beta \otimes \vert n\rangle\langle
n\vert + \alpha kv \otimes \mathbbm{1} $ on the lattice points. Here, $\vert n\rangle\langle
n\vert$ is used to make sure that $\beta$ acts only at the lattice points. The term $\alpha kv$ is valid at all positions in space (even outside the lattice points) and hence we get the term $\mathbbm{1} $. Here both $\alpha$ and $\beta$ are
$2\times 2$ Hermitian matrices.
 Now we can write the translational operator $T(x)$ (which displaces the particle from a point $x$ to $x \pm \Delta x$, $\Delta x$ being the separation between two points in space, not necessarily the separation between two lattice points) as
\begin{eqnarray}
T(x) &=& \begin{pmatrix}
\vert x - \Delta x\rangle \langle x \vert & 0 \\
0 & \vert x + \Delta x \rangle \langle x \vert
\end{pmatrix} 
\end{eqnarray}
If we denote the Hamiltonian corresponding to the translation
operator $T$ as $H_T \otimes \mathbbm{1}$, then following Appx.~\ref{append:deriv:HT1D},
\begin{eqnarray}
H_T \otimes \mathbbm{1} &=& - k v \sigma_z \,\, . 
\end{eqnarray}
In deriving the previous expression, we have taken $\hbar =1$ and have
defined $v \equiv \frac{\Delta x }{\Delta t}$ as the effective speed of the quantum particle. Throughout this article, we will always
consider $\hbar =1$. The derived Hamiltonian $H_T \otimes \mathbbm{1}$ is valid at all positions, whether the
position is a lattice point or not. 

If the Hamiltonian corresponding to the coin operation, 
$W_1(n)$, is denoted by $H_S \otimes \vert n \rangle \langle n \vert $, following Appx.~\ref{append:deriv:HS1D}, we
can write, 
\begin{eqnarray}
H_S &=& \frac{i}{\Delta \tau} \ln S \,\, .
\end{eqnarray}
Here, $\Delta \tau$ is the time interval to flip the spin, i.e. it is
the time interval the particle stays at the lattice point $n$. It is
assumed that the particle stays for a very short time-interval at the
lattice point $n$, hence $\Delta \tau$ is much smaller compared to the time scale in which the particle is translated from
one lattice point to the other. We can write the  total Hamiltonian as
\begin{eqnarray}
H &=& H_T \otimes \mathbbm{1} + \sum_n H_S \otimes \vert n \rangle \langle n \vert \,\, .
\label{Haliltonian_1D}
\end{eqnarray}
Therefore, following Appx.~\ref{append:H_x}, for any point $x$ the position space representation of the Hamiltonian is given by 
\begin{eqnarray}
H(x) &=& H_T +\sum_n H_s\delta(x-n) \,\, . \label{eq:Hamiltonian1D_position_space}
\end{eqnarray}
The Hamiltonian has a continuous term $-\sigma_z kv$ and a
Dirac comb term. This can be interpreted as Weyl Hamiltonian with a
Dirac comb potential \footnote{At a first glance, Eq.\eqref{eq:Hamiltonian1D_position_space} may look like dimensionally inconsistent but it is not, since while converting a functional from its discrete to continuous form we must divide the discrete form with a weight factor of dimension same as that of the continuous variable. In this expression we have used a weight factor of unit length.}

%%%%%%%%%%%%%%%%%%%%%%%%%%%%%%%%%%%%%%%%%%%%%%%%%%%%%
\subsection{Time evolution operator from a Dirac like Hamiltonian in 1
  D} 
\label{sec:Time evolution operator from a Dirac like Hamiltonian in 1 D}
\subsubsection{Formulation and numerical simulation} 
\label{1D generalized hamiltonian}

Let's consider the generalized Hamiltonian as derived in
Eq.\eqref{Haliltonian_1D}.  Therefore, we can write our time evolution
operator at the $n$-th lattice point as
\begin{eqnarray}
W(n,\Delta t) &=& e^{-i H_T \Delta t \otimes \mathbbm{1}}e^{-i H_S \Delta \tau \otimes \vert n \rangle \langle n \vert} 
\nonumber \\
&=& T(n) \left( S \otimes \vert n \rangle \langle n \vert \right).  \label{eq:evolution_1D}
\end{eqnarray}   
We can write down the time evolution operator here as the product of the two terms in this way because the term $\Delta t \left[ H_S, H_T
  \right] \otimes \vert n \rangle \langle n \vert$ goes to zero due to
the following reason. When the particle is at some lattice point for the time interval $\Delta \tau$, 
$\Delta x \rightarrow 0$ and $\lim_{\Delta x \to 0} H_T \Delta \tau =
0 $. Further, when the particle is not at a lattice point, $H_S =0$ by
our formulation. So we get rid of the commutation term. Physically we
can say that the particle would stay at a lattice point $n$ for an
infinitesimal time interval $\Delta \tau$. In that time interval, the effect of
$H_T$ can be neglected. Equivalently, we can say that during the
chirality flip, the translation is almost zero. Therefore the effect
of $H_T \Delta \tau$ can be neglected. The particle takes some finite time $\Delta t$ to reach the next lattice point and when it is in between two
lattice points, $H_S$ does not have any effect on its state. Hence we
can consider the action of the time evolution operator for a
particular lattice point for a finite time interval $\Delta t$, as an
operation to first mix the coin states of the particle within an infinitesimal time interval $\Delta \tau$, followed by a
translation occurring in a time interval $\Delta t - \Delta \tau$ towards a particular direction depending on its chirality. While deriving Eq.\eqref{eq:evolution_1D}, at the final step we have assumed that $\Delta x = \Delta t =1$.

In general, for a $SO(2)$ matrix $S= \begin{pmatrix}
\cos \theta & -\sin \theta \\
\sin \theta & \cos \theta 
\end{pmatrix}
$, we get the spin Hamiltonian, $H_S = \frac{\theta}{\Delta \tau} \sigma_y$. 

\begin{figure}[hbtp]
\begin{minipage}[c]{.5\linewidth}
\centering
\includegraphics[scale=0.45]{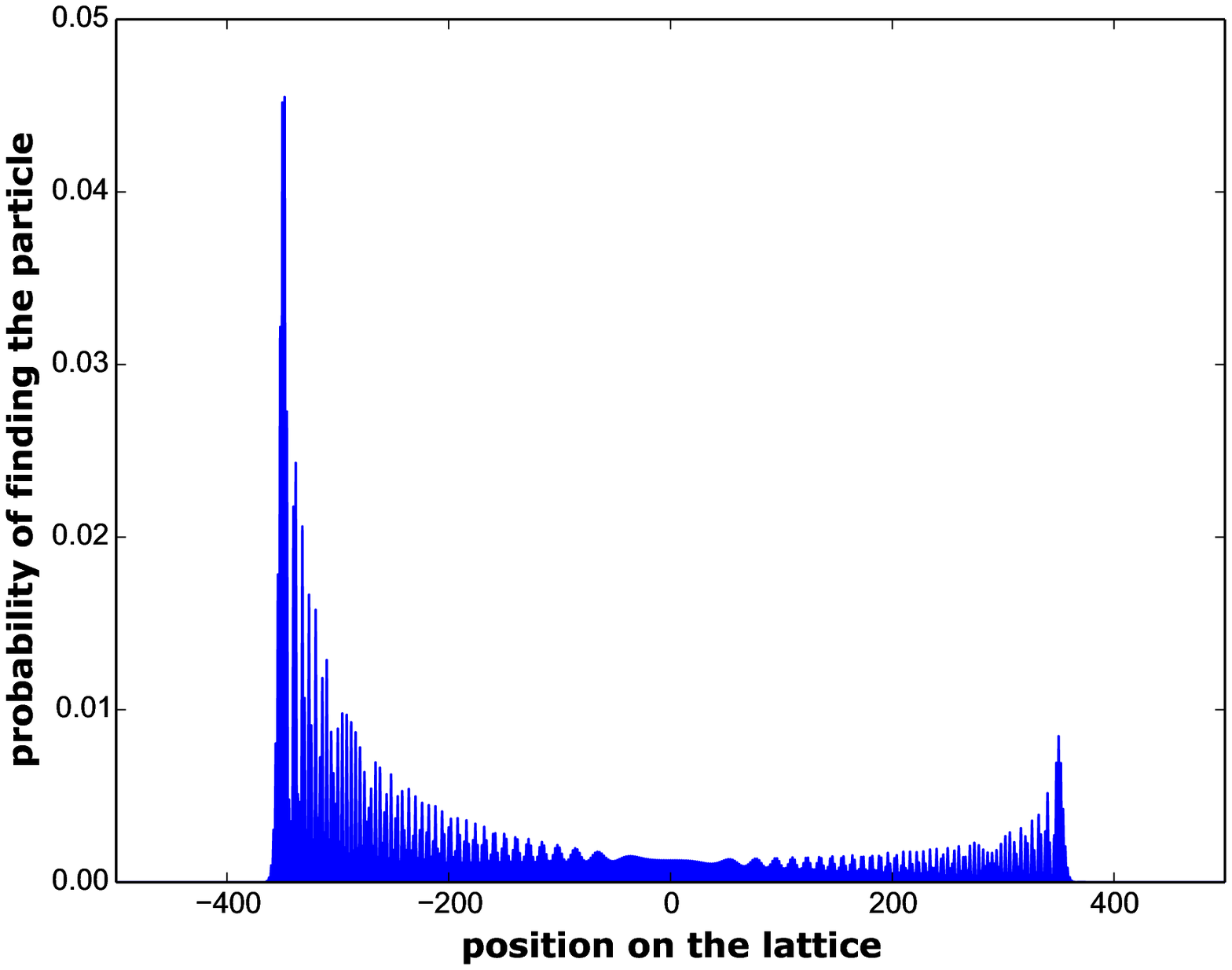}
\caption{\scriptsize In this figure the particle started quantum walk
  from the lattice point $0$. After $500$ steps probability
  distribution of the outcome is shown in the figure. }
\label{fig:1d_hada_coin}
\end{minipage}
\,\,\,\,\,\,\,\,
\begin{minipage}[c]{.5\linewidth}
\centering
\includegraphics[scale=.45]{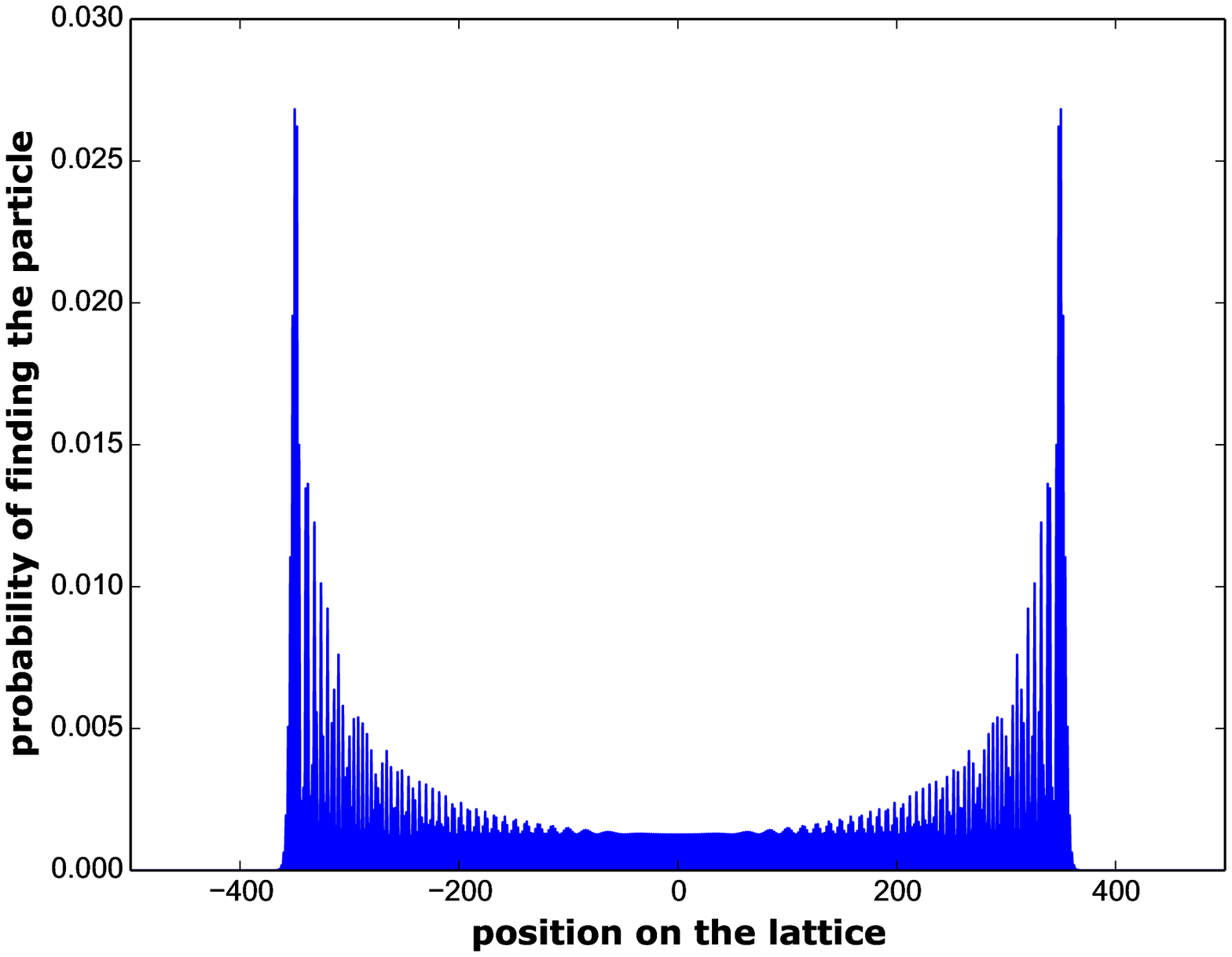}
\caption{\scriptsize In this figure the particle started quantum walk
  from the lattice point $0$. After $500$ steps probability
  distribution of the outcome is shown in the figure. }
\label{fig:1d_Y_coin}
\end{minipage}
\end{figure}

For a $SU(2)$ matrix $S=\begin{pmatrix}
\cos \theta &  i \sin  \theta \\ i \sin \theta & \cos \theta 
\end{pmatrix}$, we have $H_S= - \frac{\theta}{\Delta \tau} \sigma_x$. 
If we choose $\theta = \frac{\pi}{4}$ in the previous expression, then
we get, $S=Y=\frac{1}{\sqrt{2}}\begin{pmatrix}
1 & i \\
i & 1
\end{pmatrix}
$ and thereby get $H_S = -\frac{\pi}{4 \Delta \tau}\sigma_x$. The
result of 1-D quantum walk for this coin is shown in
Fig.\ref{fig:1d_Y_coin}. If we choose the Hadamard operator as
$S=H_c=\frac{1}{\sqrt{2}}
\begin{pmatrix}
1 & 1 \\ 1 & -1
\end{pmatrix}
$,  then the spin Hamiltonian $H_S = \frac{\pi}{2 \Delta \tau} \begin{pmatrix}
-1 + \frac{1}{\sqrt{2}} & \frac{1}{\sqrt{2}} \\
\frac{1}{\sqrt{2}} & -1 -\frac{1}{\sqrt{2}}
\end{pmatrix}   
$. The result of quantum walk with the Hadamard coin is shown in
Fig.\ref{fig:1d_hada_coin}. For both Fig.\ref{fig:1d_hada_coin} and Fig.~
\ref{fig:1d_Y_coin}, we have chosen our initial chiral state to be
$\vert \chi \rangle = \frac{1}{\sqrt{2}}(\vert + \rangle + \vert -
\rangle)$. The particle starts at the middle position of the lattice
points, so its initial position state is $\vert \xi \rangle = \vert 0
\rangle $. Therefore, the total initial state of the particle $\vert
\psi \rangle = \vert \chi \rangle \otimes \vert \xi \rangle $. The results of Fig.~\ref{fig:1d_hada_coin} and Fig.~\ref{fig:1d_Y_coin} are the standard results of one dimensional quantum walk well studied in the previous literatures. In
rest of the cases of this article, we will generally choose that the
particle starts from the middle point of the lattice structure unless
otherwise stated. So in general, only the initial chiral states will
be mentioned. Always the initial state of the particle is chosen to be a direct product
state of the coin state and the position state.

\subsubsection{Calculation of variance}

\begin{figure}[hbtp] 
 \centering
 \includegraphics[scale=.45]{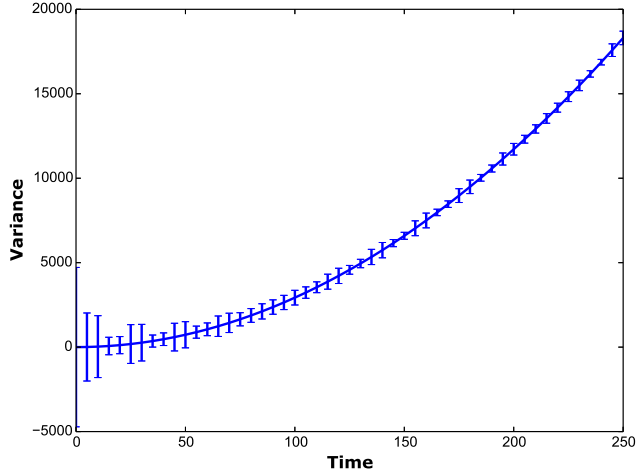}
 \caption{Plot of variance vs. time in case of 1 D quantum walk. The errorbar is with respect to the quadratic polynomial is multiplied by
   $10^4$. The Y coin has been used to generate the plot.}
 \label{variance_1d}
 \end{figure}

We calculated the variance of this 1-D walk for different no. of steps
numerically. The result shown in the Fig.\ref{variance_1d} is for the
Y coin and an initial chiral state $\vert \chi \rangle =
\frac{1}{\sqrt{2}}(\vert + \rangle + \vert - \rangle)$. The plot can
be approximated as a polynomial of the form $ 0.47175578 + 0.00091395
t + 0.29289026 t^2 $. Clearly the leading order term in the variance $\sim t^2$, unlike the
classical walk as already pointed out in Ref.~\citep{Kempe_2003} and others.

\vspace{4 cm}

\subsection{Two Dimensional case for a square lattice}
\label{sec : qrw_square_lattice}
In Ref.~\citep{Chandrashekar_2013}, while formulating random walk
on a square lattice, the Hamiltonian has been taken in additive form
and the total time evolution operator has been taken as the product of the evolution
operator acting along each axis. It implies that the particle has to
move in $x-y$ or $y-x$ order only. It does not have the choice to move
along the same axis in two consecutive steps. This approach can be
thought of taking two steps (in $x-y$ or $y-x$ order only) in each operation of the evolution operator. So we call it two step
approach in the following sections of this article.  But in our approach, the
particle is allowed to move along any axis in each step. Hence in this
single step approach there is no forbidden path for the walk.

\subsubsection{Formulation} \label{2D_square_hamiltonian}
In random walk, a particle can move along any of the paths available to it from a lattice point. So in case of quantum walk the dimensionality of
the coin (i.e. the dimension of the Hilbert space on which the coin acts) is determined by the number of paths the particle can choose. In
all dimensions higher than one, we have two types of chirality of the
particle for the approach described in this article. One is to choose which axis the particle would move along in
the next step and the other to decide whether the particle would go to
the positive or to the negative direction of the chosen axis.

In case of a square lattice, the particle has two choice of axes at
each lattice point. In this case, analogous to
Eq.~\eqref{eq:Hamiltonian1D_position_space}, we can say that the Hamiltonian at a point
$(x, y)$ for 2-D walk will have the form
\begin{eqnarray}
H(x,y) = \sum_{n,m} H_s \delta(x-n)\delta(y-m) + H_T \otimes \mathbbm{1} \,\, .
\end{eqnarray}		
Here, 
\begin{eqnarray}
H_{S} = \frac{i}{\Delta \tau} \ln \left( S_x \otimes M_1 + S_y \otimes M_2 \right) \, , \\
H_T = - \left( k_x v_x \begin{pmatrix}
1 & 0 \\ 0 & 0 \end{pmatrix} + k_y v_y \begin{pmatrix}
0 & 0 \\ 0 & 1 \end{pmatrix}  \right) \otimes \sigma_z \, . \end{eqnarray}
Where $S_{x}$ and $S_{y}$ are given as, 
\begin{eqnarray}
S_{x} = \begin{pmatrix}
 S_{11} & S_{12} \\ 0 & 0
\end{pmatrix} \,\,\,\,\,\,\ {\rm and} \,\,\,\,\,\,\ S_{y} = \begin{pmatrix}
 0 & 0 \\ S_{21} & S_{22}
\end{pmatrix} .
\end{eqnarray}
We define $S = S_{x} + S_{y}$ as the chirality operator for choosing the $X$ or the $Y$ axis. $M_1$ and $M_2$ are chirality operators for choosing positive or negative direction of a particular axis. Now, if we choose discrete set of lattice points such that the position eigenstate is $\vert n,m \rangle$, with the condition $\langle n,m\vert p,q \rangle = \delta_{np} \delta_{mq}$ ($n,m,p,q$ are integers), then following similar approach as shown in Sec. \ref{1D generalized hamiltonian}, we can write (Appx.~\ref{append:deriv:W2D}),
 \begin{eqnarray}
 W_{nm} &=& T_{nm} \left( \left( S_x \otimes M_1  + S_y \otimes M_2 \right) \otimes \vert n,m \rangle \langle n,m \vert \right) , \label{evolution_additive}
 \end{eqnarray} 
where \begin{eqnarray} 
T_{nm} = \begin{pmatrix}
\vert n-1,m \rangle \langle n,m \vert & 0 & 0 & 0 \\
0 & \vert n+1,m \rangle \langle n,m \vert & 0 & 0 \\
0 & 0 & \vert n,m-1 \rangle \langle n,m \vert & 0 \\
0 & 0 & 0 & \vert n,m+1 \rangle \langle n,m \vert
\end{pmatrix} . \nonumber
\end{eqnarray}
is the translation operator. The result of
Eq.\eqref{evolution_additive} shows that the time evolution operator
is additive in nature. This clearly differs from the two-step \footnote{ In the two-step approach if the particle moves along $x$ axis at
  one step, the next step must be along $y$-axis. So we can interpret
  the method as follows: first $W_x$ operates on the particle and
  displaces it along $x$-direction. When it reaches to the next
  lattice point, $W_y$ operates. So in this case the time evolution
  operator is different at different lattice points. The total time
  evolution operator for the lattice is the sum of the time evolution
  operator at all the lattice points. If the particle starts to move
  along $X$ direction from the point $(0,0)$, at all the even points
  (points for which $x+y$ is even) the operator $W_x$ will act while
  at all the odd points (points for which $x+y$ is odd) the operator
  $W_y$ will act. Although the operator is not actually
  multiplicative, the net effect in two-step can be described by
  $W_yW_x$ because after evolution by $W_x$ the particle would reach
  next lattice point where the evolution is governed by $W_y$. This
  can happen if at the points where $W_x$ is operating, $k_yv_y = 0$
  and at the points where $W_y$ is operating, $k_xv_x = 0$. In this way, we can  reproduce the two-step approach of Ref.~\citep{Chandrashekar_2013} starting from our Hamiltonian formulation of single-step walk.} approach of Ref.~\citep{Chandrashekar_2013} where it is assumed that the two
dimensional time evolution operator is the product of one dimensional
time evolution operator in the $X$ and $Y$ direction.  Now for
the total chirality operator $\mathbb{M} = S_x \otimes M_1 + S_y
\otimes M_2$ we can use different types of mixing operators,
e.g. Grover, DFT or the higher dimensional Hadamard coin as discussed
in the next section.

%%%%%%%%%%%%%%%%%%%%%%%%%%%%%
\subsubsection{Plots for different coins} 
\label{sec: plot_square_lattice}
\paragraph{Case 1 :} First let's choose the coin as the direct product of two Hadamard coins.

\begin{figure}[hbtp]
\begin{minipage}[c]{.5\linewidth}
\centering
\includegraphics[scale=.27]{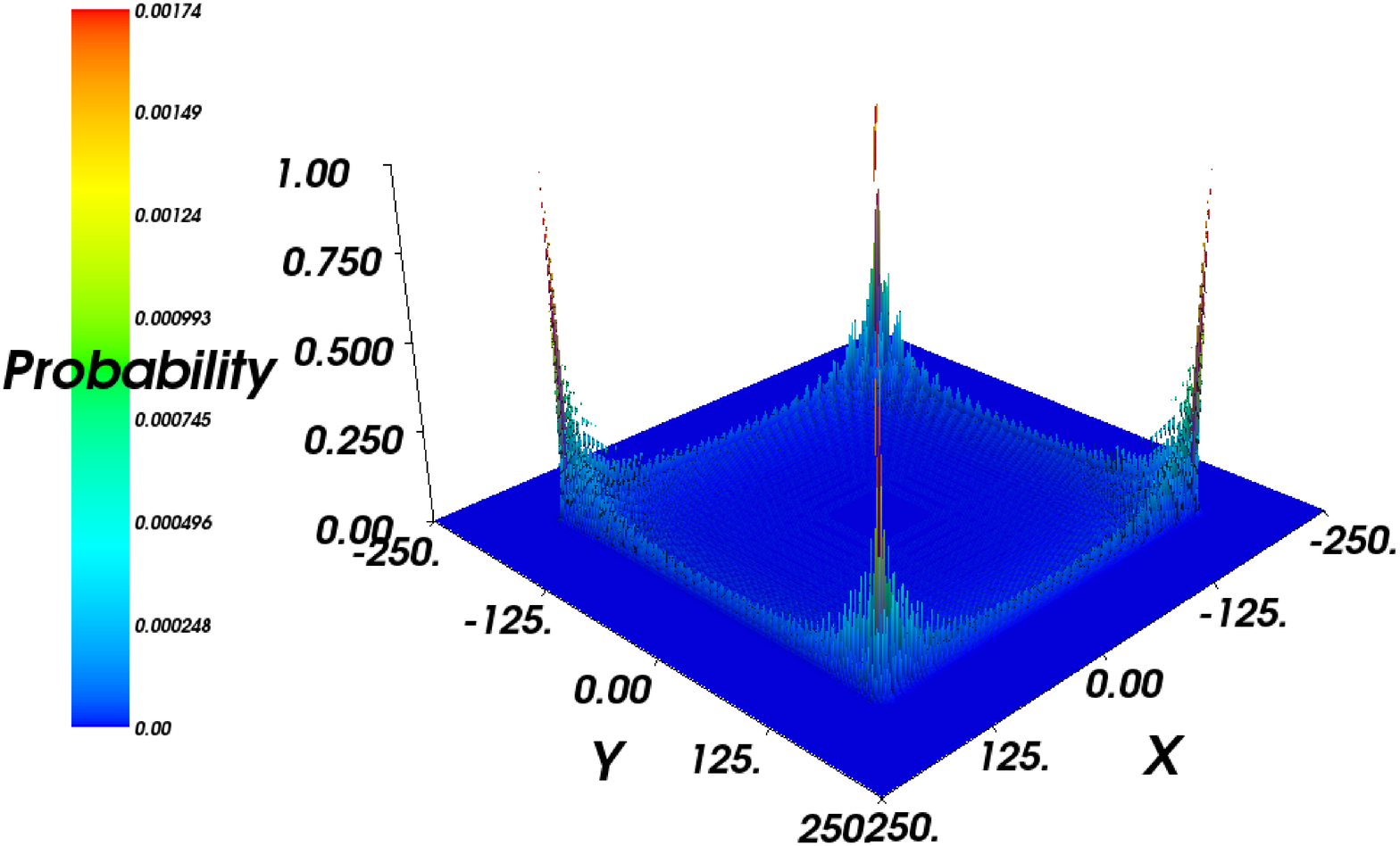} %\vspace{.2 cm}
\caption{Probability of finding the particle after 250 steps starting from the origin. This plot is with the two-step approach using the coin of Eq.\eqref{eq:hada_hada_coin}.}
\label{two_step_hada_hada}
\end{minipage}
\,\,\,\,\,\,
\begin{minipage}[c]{.5\linewidth}
\centering
\includegraphics[scale=.25]{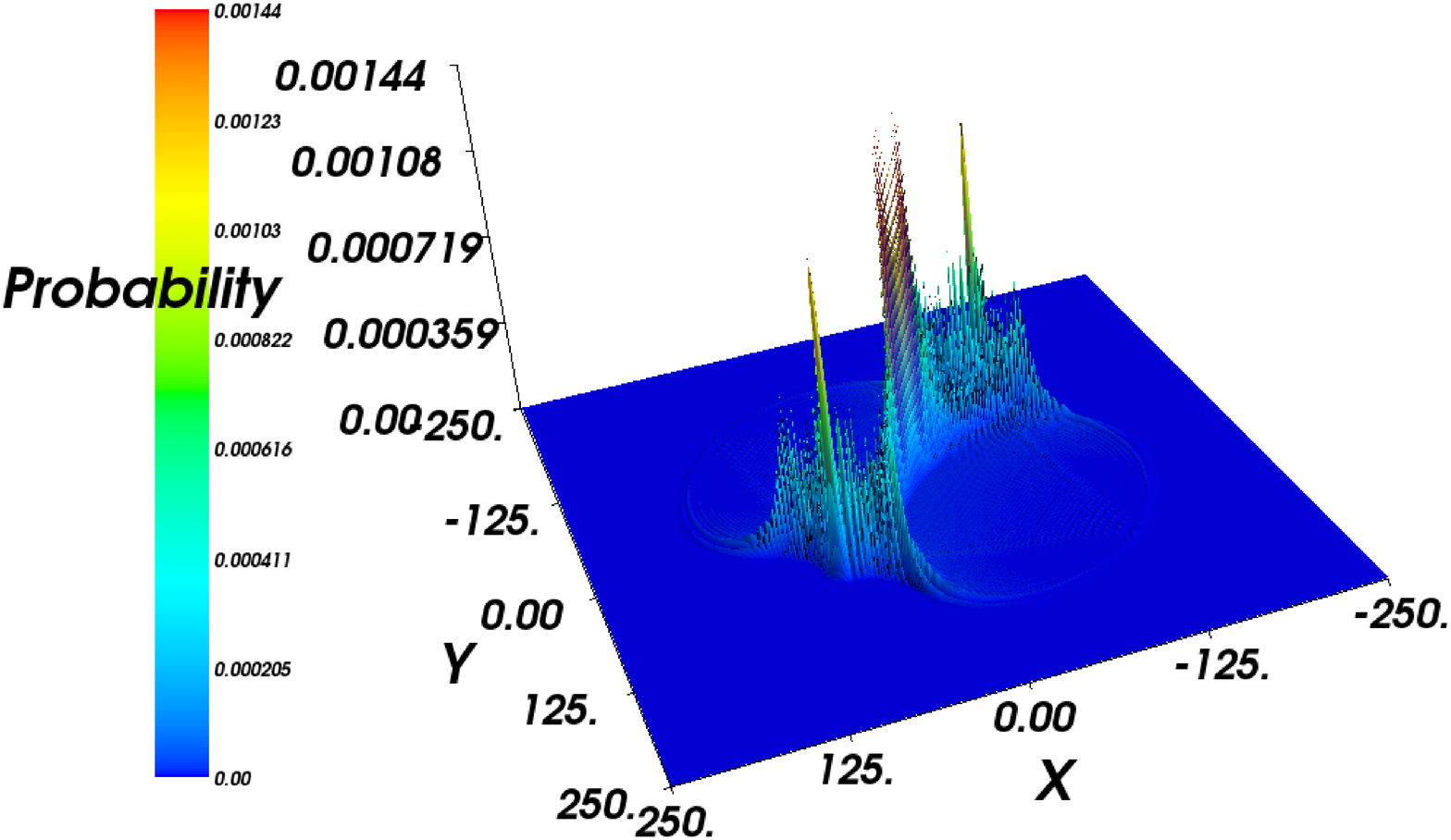} 
\caption{Probability of finding the particle after 250 steps starting from the origin. This plot is with the one-step approach using the coin of Eq.\eqref{eq:hada_hada_coin}}
\label{single_step_haha_hada}
\end{minipage}
\end{figure}

In that case, 
\begin{eqnarray}
\mathbb{M} = \frac{1}{2} 
\begin{pmatrix}
1 & 1 & 1 & 1 \\
1 & -1 & 1 & -1 \\
1 & 1 & -1 & -1 \\
1 & -1 & -1 & 1 
\end{pmatrix} . \label{eq:hada_hada_coin}
\end{eqnarray}

We choose our initial chiral state to be $\vert \chi \rangle = \frac{1}{\sqrt{2}}(\vert 1 \rangle + i \vert 2 \rangle) \otimes \frac{1}{\sqrt{2}}(\vert + \rangle + i \vert - \rangle )$ . From now onwards $\vert 1 \rangle , \vert 2 \rangle$ will denote the chiral state of the particle to choose the $X$ and the $Y$ axis respectively and $\vert + \rangle, \vert - \rangle$ denote whether the particle will move along the positive or negative side  of a chosen axis respectively. The result of the walk governed by this coin is shown in Fig.\ref{two_step_hada_hada} and Fig.~\ref{single_step_haha_hada}. The spin Hamiltonian for  this coin is given as 
\begin{eqnarray}
H_S = \frac{\pi}{4 \Delta \tau} \begin{pmatrix}
-1 & 1 & 1 & 1 \\ 1 & -3 & 1 & -1 \\ 1 & 1 & -3 & -1 \\ 1 & -1 & -1 & -1
\end{pmatrix} .
\end{eqnarray}

\paragraph{Case 2 :} Now we choose the Grover coin described in Ref.~\citep{Kempe_2003}. Therefore,
\begin{eqnarray}
\mathbb{M} &=& \frac{1}{2} 
\begin{pmatrix}
-1 & 1 & 1 & 1 \\
1 & -1 & 1 & 1 \\
1 & 1 & -1 & 1 \\
1 & 1 & 1 & -1 
\end{pmatrix} , \label{eq:grover_coin}
\end{eqnarray}

\begin{figure}[hbtp]
\begin{minipage}[c]{.5\linewidth}
\centering
\includegraphics[scale=.26]{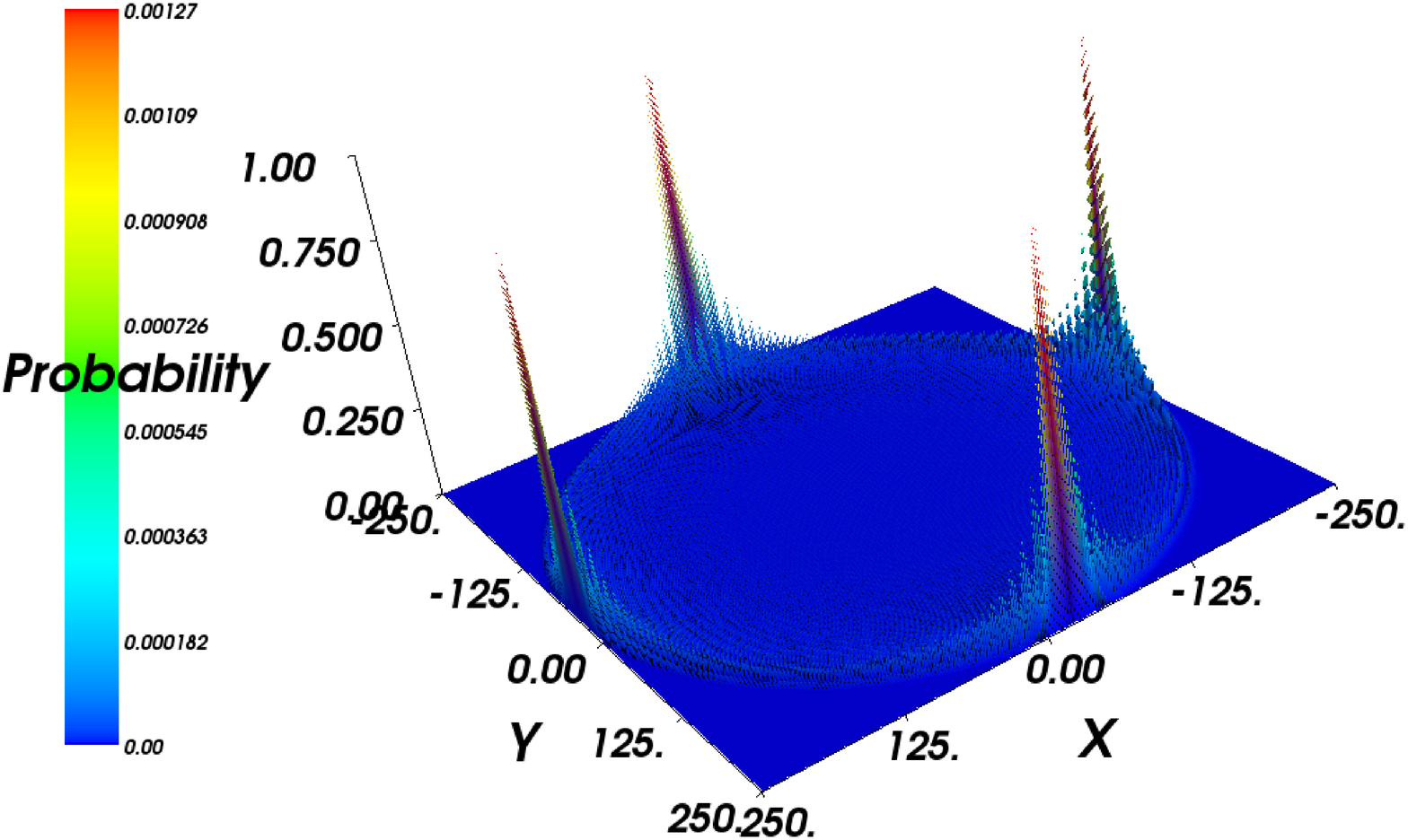}
\caption{Probability of finding the particle after 250 steps starting from the origin. This plot is with the two-step approach using the Grover coin of Eq.\eqref{eq:grover_coin}.}
\label{two_step_grover}
\end{minipage}
\,\,\,\,\,\,
\begin{minipage}[c]{.5\linewidth}
\centering
\includegraphics[scale=.26]{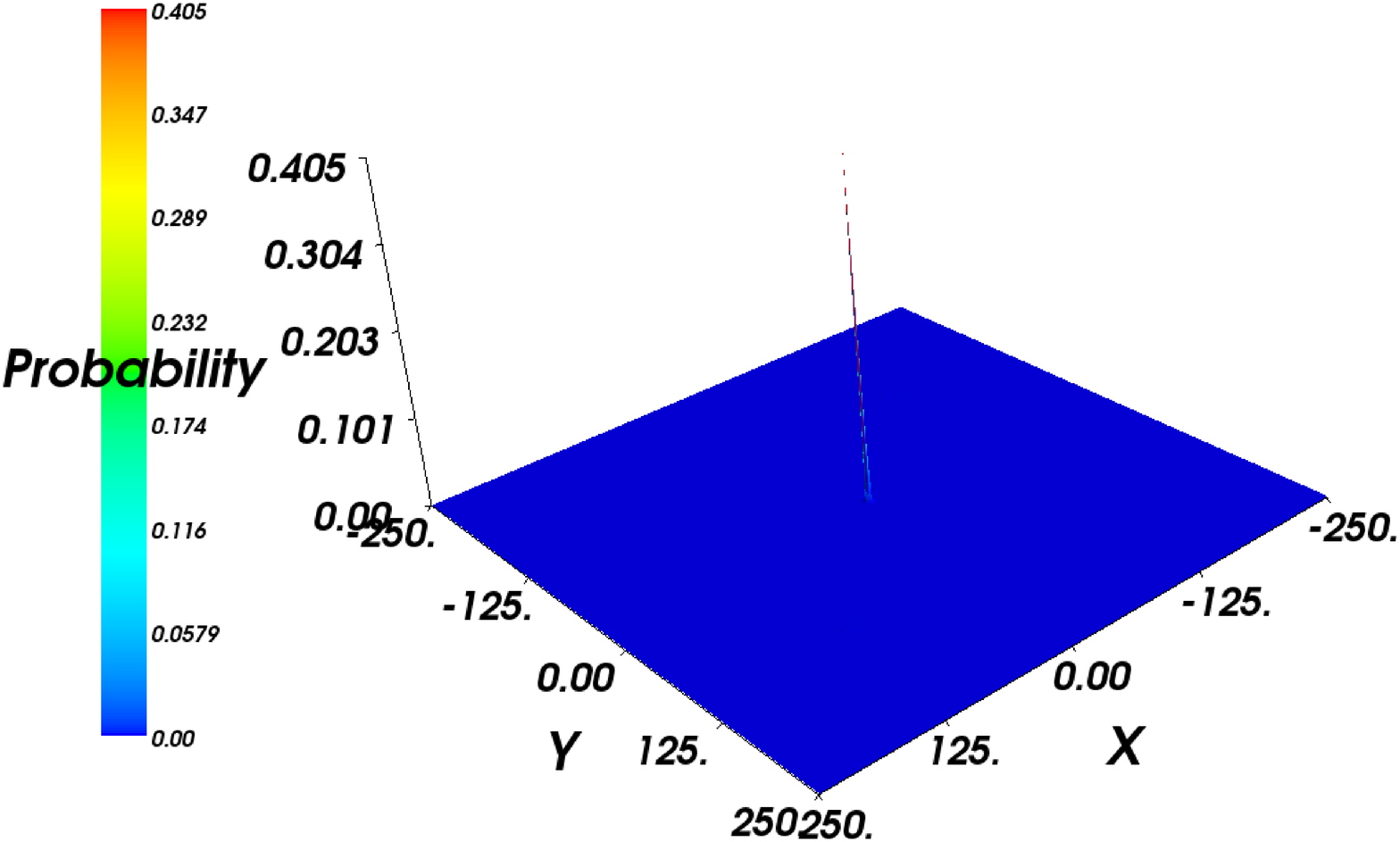}
\caption{Probability of finding the particle after 250 steps starting from the origin. This plot is with the one-step approach using the Grover coin of Eq.\eqref{eq:grover_coin}.}
\label{single_step_grover}
\end{minipage}
\end{figure}

and we choose our initial chiral state as $\vert \chi \rangle = \frac{1}{\sqrt{2}}(\vert 1 \rangle - \vert 2 \rangle) \otimes \frac{1}{\sqrt{2}}(\vert + \rangle - \vert - \rangle )$. The result is as shown in Fig.~\ref{two_step_grover} and Fig.~\ref{single_step_grover}. The spin Hamiltonian $H_S$ for this coin is given by,
\begin{eqnarray}
H_S = \frac{\pi}{4 \Delta \tau} \begin{pmatrix}
-3 & 1 & 1 & 1 \\ 1 & -3 & 1 & 1 \\ 1 & 1 & -3 & 1 \\ 1 & 1 &  1 & -3
\end{pmatrix} .
\end{eqnarray}

\paragraph{Case 3 :} Now we choose the DFT coin described in Ref.~\citep{Kempe_2003}. Therefore,

\begin{figure}[hbtp]
\begin{minipage}[c]{.5\linewidth}
\centering
\includegraphics[scale=.25]{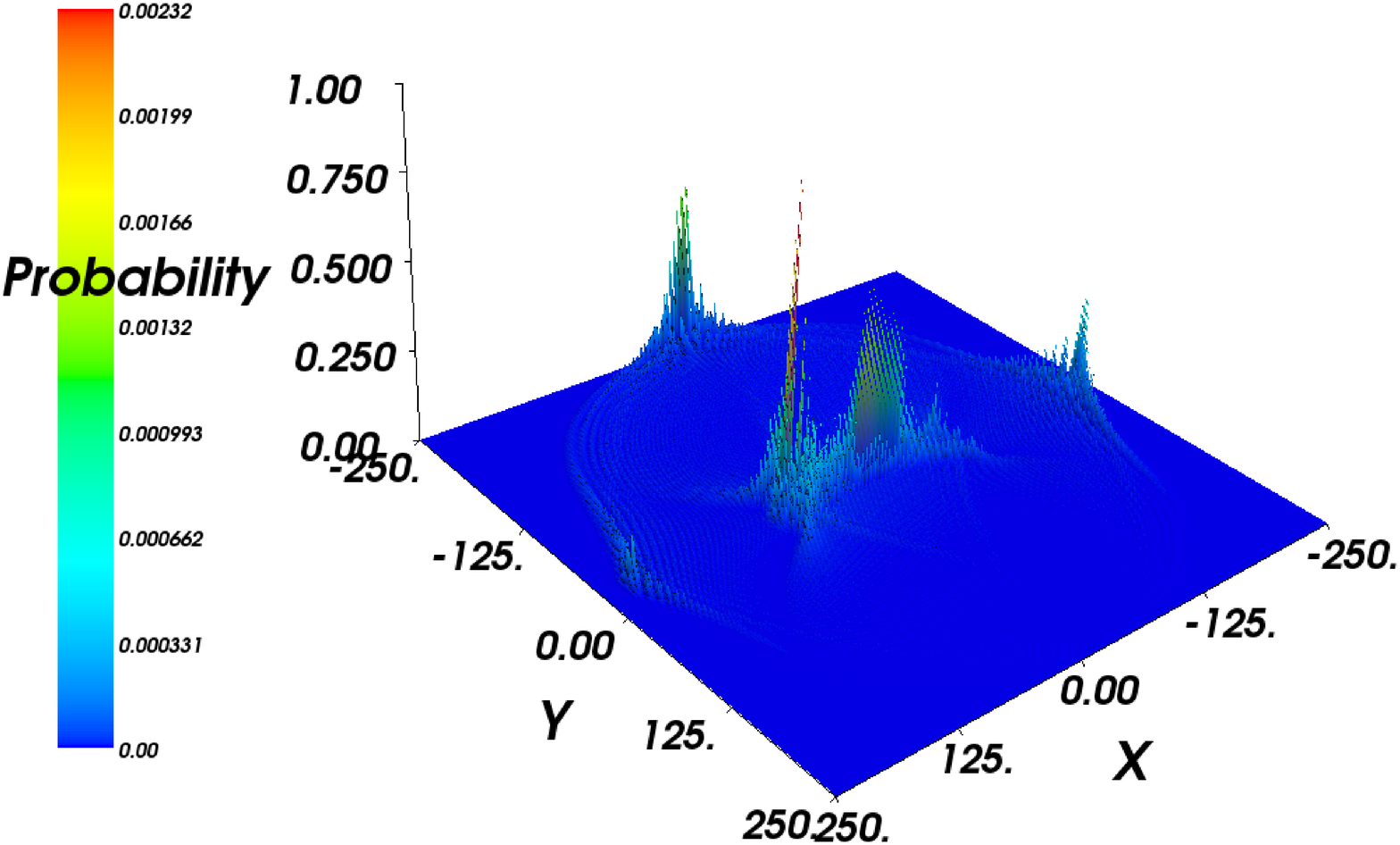}
\caption{Probability of finding the particle after 250 steps starting from the origin. This plot is with the two-step approach using the DFT coin of Eq.\eqref{eq:DFT_coin}.}
\label{two_step_DFT}
\end{minipage}
\,\,\,\,\,\,
\begin{minipage}[c]{.5\linewidth}
\centering
\includegraphics[scale=.25]{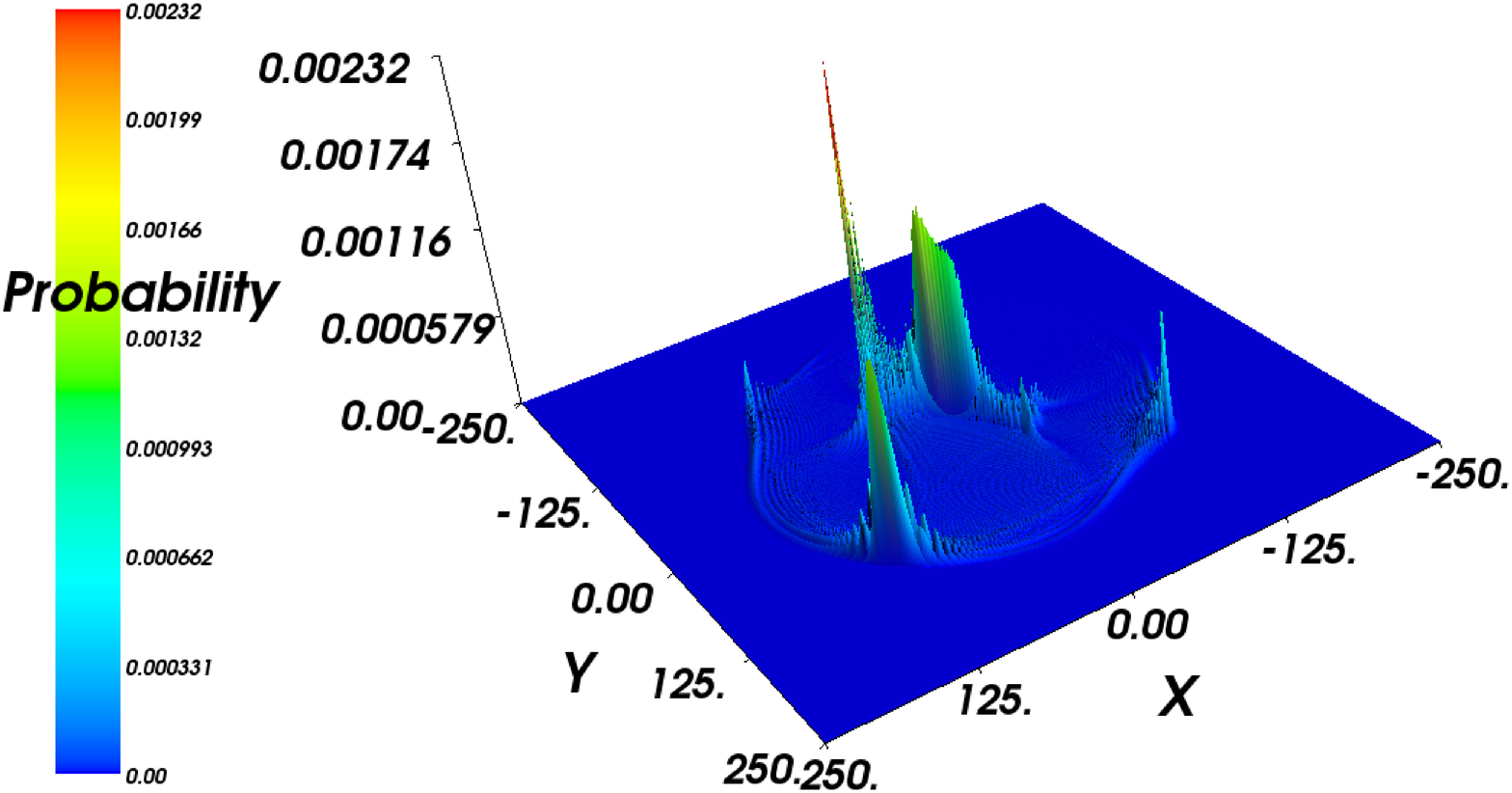}
\caption{Probability of finding the particle after 250 steps starting from the origin. This plot is with the one-step approach using the DFT coin of Eq.\eqref{eq:DFT_coin}.}
\label{single_step_DFT}
\end{minipage}
\end{figure}

\begin{eqnarray}
\mathbb{M} &=& \frac{1}{2} 
\begin{pmatrix}
1 & 1 & 1 & 1 \\
1 & i & -1 & -i \\
1 & -1 & 1 & -1 \\
1 & - i & -1 & i 
\end{pmatrix}  \,\,\,\,\,\, ({\rm\, where,}\,\, i = \sqrt{-1} \,) , \label{eq:DFT_coin}
\end{eqnarray}

and we choose our initial chiral state $\vert \chi \rangle = \frac{1}{\sqrt{2}}(\vert 1 \rangle - \vert 2 \rangle) \otimes \frac{1}{\sqrt{2}}(\vert + \rangle - \vert - \rangle )$. The result is as shown in Fig.\ref{two_step_DFT} and \ref{single_step_DFT}. The spin Hamiltonian for this coin is given by 
\begin{eqnarray}
H_S = \frac{\pi}{4 \Delta \tau} \begin{pmatrix}
-1 & 1 & 1 & 1 \\ 1 & -2 & -1 & 0 \\
1 & -1 & -1 & -1 \\ 1 & 0 & -1 & -2
\end{pmatrix} .
\end{eqnarray}

The plots of Fig.~\ref{two_step_hada_hada}, \ref{two_step_grover} and \ref{two_step_DFT} are the standard two-step quantum walks well studied before in the other literatures but they have been shown here to compare the results with that of the single-step approach of the corresponding coins.

%%%%%%%%%%%%%%%
\subsubsection{Calculation of variance}
\begin{figure}[hbtp]
\centering
\includegraphics[scale=.5]{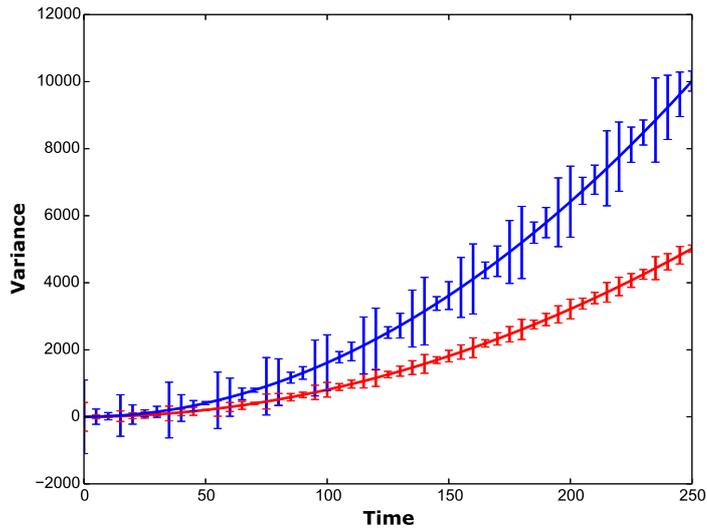}
\caption{Plot of variance with time in case of a square lattice. The red curve shows variance in case of the single step approach and the blue one for two step approach. The errorbar with respect to the quadratic polynomial in both cases is multiplied by a factor of $10^3$. The DFT coin of Eq.\eqref{eq:DFT_coin} has been used to generate the plot.} \label{plot_variance_square_combined}
\end{figure}

We calculated the variance of the probability distribution for the square lattice case and result has almost similar form with all types of coins we have used. So we mention here only the result obtained
with the DFT coin. The above figure is obtained with an initial chiral state, $\vert \chi \rangle = \frac{1}{\sqrt{2}}(\vert 1 \rangle
+ \vert 2 \rangle) \otimes \frac{1}{\sqrt{2}}(\vert + \rangle + \vert
- \rangle) $. The red color plot in
Fig.\ref{plot_variance_square_combined} is for single step approach
and can be approximated with an approximated polynomial of the form $
0.42904373 + 0.08807425 t + 0.07993539 t^2$ and the blue color plot is
for two step approach and can be described with an approximated
polynomial $ 1.09843965 + 0.17597539 t + 0.15954993 t^2 $. So we see
that the variance of the probability distribution in QRW has quadratic
nature with time in case of 2D square lattice, similar to the form for
1D case as mentioned in Ref.~\citep{Nayak_Vishwanath_2000}, \citep{Kempe_2003}.

%%%%%%%%%%%%%%%%%%%%%%%%%

\section{Quantum walk on Graphene Lattice}
\label{sec: qrw_graphene} 
 The Graphene Hamiltonian according to Ref.~\citep{Castro_Neto_2009}  is given by $H=v_F\vec{\sigma}.\vec{p}$ which is valid only at positions very close to a lattice point. Taking $\hbar v_F=1$, we can write $H=\vec{\sigma}.\vec{k}=\sigma_x k_x +\sigma_y k_y$. We will see that this Hamiltonian can generate an infinitesimal time evolution operator close to a lattice point. To generate a finite time evolution operator we need the Hamiltonian for translation operator to act at positions in between two lattice points.

\subsection{Formulation of the problem} 
\begin{figure}[hbtp]
\centering
\includegraphics[scale=.5]{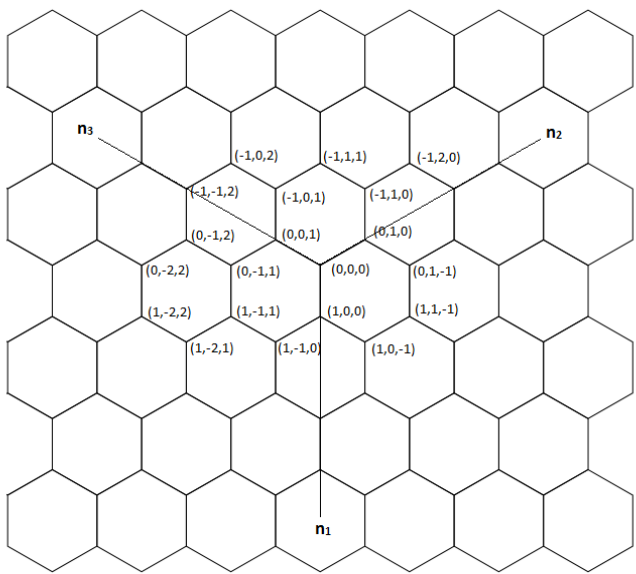}
\caption{Here Graphene lattice is presented with the three axes $n_1$, $n_2$ and $n_3$. Some of the lattice points are specified on the lattice.}
\label{graphene_lattice}
\end{figure}

To describe random walk problem on Graphene lattice, we have to consider the three available paths at each point. So we need three axes as defined by $n_1$, $n_2$ and $n_3$ as shown in the Fig.\ref{graphene_lattice}. We denote the directions along these three axes by 1, 2 and 3 respectively. Clearly these three quantities $n_1, n_2, n_3$ are not linearly independent but they are dependent by the relations 
\begin{eqnarray}
x = \frac{\sqrt{3}(n_2 - n_3)}{2} \,\,\,\, {\rm and} \,\,\,\, y = - n_1 + \frac{(n_2 + n_3)}{2} \,\, .
\end{eqnarray} 	
By choosing these three axes we can easily extend our previous approach for square lattice in this case. In this approach, the position ket at each lattice point $( x,y )$ is denoted by $ \vert n_1, n_2, n_3 \rangle $. 	
For Graphene lattice, we can classify the lattice points into two
types. The points for which the sum of $n_1$, $n_2$ and $n_3$ is even,
are called even points and those having an odd sum of lattice indices
are classified as odd points. Let us consider the two nearby lattice
points, say the even point $(0,0,0)$ denoted by point $1$ and the odd
point $(0,1,0)$ denoted by point $2$. From the point $1$, the particle
can move only to the positive direction of all the three axes (see
Fig.\ref{graphene_lattice}). But from the point $2$, the particle
moves only along the negative direction of all the three
axes. Clearly, the geometry itself constraints the direction of
movement here. We can describe this by using two kets as $\vert
+\rangle$ and $\vert-\rangle$ which denotes the chirality of going
along the negative or positive direction respectively as in the
earlier cases. The only difference in this case is that the particle
cannot be at superposition of $\vert+\rangle$ and $\vert-\rangle$, it
must be at either of them. It is clear from the geometry that if the
particle is in $\vert\pm\rangle$ state at some step, at the next step
it must be in $\vert\mp\rangle$ state. For the lattice as described in
Fig. \ref{graphene_lattice}, for even points, the particle is always
in $\vert - \rangle$ state and for the odd points it is always in
$\vert + \rangle$ state as per our convention. Clearly the operator
needed to describe this kind of motion is given by $\begin{pmatrix}
  0&a\\b&0 \end{pmatrix}$. Again, we need a $3\times 3$ matrix to make
the superposition of three chiral states which determine the choice
one of the three axes. After choosing the chirality, the state must
translate. Following Eq.\eqref{evolution_additive}, we can write the
general time evolution operator as $W(n_1,n_2,n_3) =
T(n_1,n_2,n_3)\left( \left( S_1 \otimes M_1 +S_2 \otimes M_2 + S_3
\otimes M_3 \right) \otimes \vert n_1,n_2,n_3\rangle \langle
n_1,n_2,n_3 \vert \right)$. Here $S_1, S_2$ and $S_3$ are $3 \times 3$
matrices which have only the $1^{st}$, $2^{nd}$ and $3^{rd}$ rows
non-zero respectively. We can choose any three dimensional unitary
operator as $S$ (where, $S = S_1 + S_2 + S_3)$. We would use the
Graphene Hamiltonian to find out $M_1, M_2$ and $M_3$. We can express
$k_x$, $k_y$ the components of the wave vector $\vec{k}$, in terms of
$k_1$, $k_2$ and $k_3$ in the following way,
\begin{eqnarray}
 k_x = \frac{\sqrt{3}}{2}\left( k_2 - k_3\right) \,\,\,\,  {\rm and} \,\,\,\, k_y = - k_1 + \frac{1}{2}\left( k_2 + k_3 \right) \,\, .  
\end{eqnarray} 
So close to a lattice point, the Graphene Hamiltonian  can be written as
 \begin{eqnarray}
H_{xy} &=& \left( k_x \sigma_{x} + k_y \sigma_{y} \right) \otimes \vert x,y \rangle \langle x,y \vert \nonumber \\
&=& \left( \begin{pmatrix}
 0 & ik_1\\-ik_1 & 0
 \end{pmatrix} + 
 \begin{pmatrix}
 0 & i \omega^2 k_2\\
 -i\omega k_2 & 0
 \end{pmatrix} + 
 \begin{pmatrix}
 0 & i\omega k_3 \\ -i\omega^2 k_3 & 0
 \end{pmatrix}\right) \otimes \vert x,y\rangle\langle x,y\vert \,\, .  \label{eq:graphene_hamiltonian}
 \end{eqnarray}
where, $\omega = \frac{-1+\sqrt{3} i}{2},\,\, i = \sqrt{-1} $.  Since
the above form of the Hamiltonian is valid close to the lattice point,
i.e. within a very small region around a lattice point, it would represent infinitesimal time
evolution of the quantum particle near the lattice point. This
Hamiltonian shifts the chirality of the particle and translates it
infinitesimally. The Hamiltonian for a position away from the lattice
point is given by $H_T$, the translational Hamiltonian. So first the
Graphene Hamiltonian acts and leave the particle slightly away from
the lattice point. After that the translation operator takes it to
the next lattice point. Now following Eq.\eqref{eq:graphene_hamiltonian} and neglecting the subscript in $H_{xy}$, we can write,
 \begin{eqnarray}
-i H &=& -i\left[\begin{pmatrix}
 0 & i k_1\\-i k_1& 0
 \end{pmatrix} +\begin{pmatrix}
 0&i\omega^2 k_2\\-i\omega k_2&0
 \end{pmatrix}+\begin{pmatrix}
 0&i\omega k_3\\-i\omega^2 k_3&0
 \end{pmatrix}  \right]\otimes \vert n_1,n_2,n_3\rangle\langle
n_1,n_2,n_3\vert\nonumber \\
&=& -i\left[\begin{pmatrix}
0 & 1 + i k_1 \\ 1-ik_1 & 0
\end{pmatrix} + \begin{pmatrix}
0 & \omega^2(1+ik_2)\\ \omega(1-ik_2) & 0
\end{pmatrix}+\begin{pmatrix}
0 & \omega(1+ik_3) \\ \omega^2(1-ik_3) & 0
\end{pmatrix} \right. \nonumber \\
&& \left. + i\begin{pmatrix}
 0&1+\omega+\omega^2 \\ 1+\omega+\omega^2 & 0
\end{pmatrix} \right]\otimes \vert n_1,n_2,n_3\rangle\langle
n_1,n_2,n_3
\vert \,\, . \nonumber \\ 
\Rightarrow \mathcal{I} -iH\Delta t &=& -i \left[\begin{pmatrix}
0 & 1+ik_1\Delta x_1 \\ 1-ik_1\Delta x_1 & 0
\end{pmatrix}+\begin{pmatrix}
0 &\omega^2(1+ik_2\Delta x_2)\\ \omega(1-ik_2\Delta x_2) & 0
\end{pmatrix} \right. \nonumber \\
&& \left. + \begin{pmatrix}
0&\omega(1+ik_3\Delta x_3)\\\omega^2(1-ik_3\Delta x_3) & 0 
\end{pmatrix}\right]\otimes \vert n_1,n_2,n_3\rangle\langle
n_1,n_2,n_3
\vert+I \,\, .\nonumber
\end{eqnarray}
Neglecting the identity operator above (which is not involved in any
physical operations) we get three infinitesimal time evolution
operators as
\begin{eqnarray}
W_1(n_1,n_2,n_3) &=& 
\begin{pmatrix}
0 & \vert n_1 - \Delta x_1, n_2, n_3 \rangle \langle n_1, n_2, n_3 \vert \\
\vert n_1 + \Delta x_1, n_2, n_3 \rangle \langle n_1, n_2, n_3 \vert & 0
\end{pmatrix} \,\, , \nonumber \\
W_2(n_1,n_2,n_3) &=& 
\begin{pmatrix}
0 & \omega^2 \vert n_1 , n_2 - \Delta x_2, n_3 \rangle \langle n_1,
n_2, n_3 
\vert \\
\omega \vert n_1 , n_2 + \Delta x_2, n_3 \rangle \langle n_1, n_2, n_3 \vert & 0
\end{pmatrix} \,\, , \nonumber \\
W_3(n_1,n_2,n_3) &=& 
\begin{pmatrix}
0 & \omega \vert n_1 , n_2, n_3 - \Delta x_3 \rangle \langle n_1, n_2,
n_3 
\vert \\
\omega^2 \vert n_1, n_2, n_3 + \Delta x_3\rangle \langle n_1, n_2, n_3 \vert & 0
\end{pmatrix} \,\, . \nonumber 
\end{eqnarray}
In the previous equations we have assumed that the $-i$ can be
absorbed in the normalization of the initial chiral state of the
particle. More over the $\Delta x_i$'s which appear are assumed to be
very small displacements around the lattice points, much smaller than
the lattice spacings. As the particle is $\Delta x_i$ distance away
from a typical lattice point the Hamiltonian is governed by the term
$H_T$, which is obtained from the translational operator $T^\prime
(n_1 \pm 1, n_1 \pm \Delta x_1; n_2 \pm 1, n_2 \pm \Delta x_2; n_3 \pm
1, n_3 \pm \Delta x_3)$, responsible for translating the
particle from the very neighbourhood of a lattice point to the adjacent one. When $T^\prime$ operates
on $ \left(S_1 \otimes W_1 + S_2 \otimes W_2 + S_3 \otimes
W_3\right)$, we can write the total time evolution operator in terms
of the of an effective translation operator $T\left(n_1, n_2, n_3\right)$ as
\begin{eqnarray}
W\left(n_1, n_2, n_3\right) &=& T\left(n_1, n_2, n_3\right) 
\left( \left( S_1 \otimes M_1 +S_2 \otimes M_2 + S_3 \otimes M_3
\right) \otimes \vert n_1,n_2,n_3\rangle \langle n_1,n_2,n_3 \vert \right)
\end{eqnarray} 
where, 
\begin{eqnarray}
M_1 = \begin{pmatrix}
0 & 1 \\ 1 & 0
\end{pmatrix} \,\, , \,\,
M_2 = \begin{pmatrix}
0 & \omega^{2} \\ \omega & 0
\end{pmatrix} \,\, {\rm and} \,\,
M_3 = \begin{pmatrix}
0 & \omega \\ \omega^{2} & 0
\end{pmatrix} \nonumber
\end{eqnarray}
and $T\left( n_1, n_2, n_3\right)$ is the effective translational operator 
defined as, 
\begin{eqnarray}
T(n_1,n_2,n_3) &=& {\rm diag} \left(
\vert n_1 - 1, n_2, n_3 \rangle , \,\,  \vert n_1 + 1, n_2, n_3 
\rangle , \,\,  \vert n_1, n_2 - 1, n_3 \rangle , \,\,  \vert n_1, n_2
+ 1, n_3 \rangle , \,\, \vert n_1, n_2, n_3 - 1 \rangle, \right. \nonumber \\
&& \left.  \vert n_1, n_2, n_3 + 1 \rangle \right)  \langle n_1, n_2,
n_3 \vert\,.   
\nonumber
\end{eqnarray}
Here the effective translation operator $T$ is a combination of two
translations \footnote{The translational operator is generated by two
  translational operators as follows:
\begin{eqnarray}
T(n_1,n_2,n_3) = T^\prime\left( n_1\pm 1, n_1 \pm \Delta x_1 ;  n_2
\pm 1, n_2 \pm \Delta x_2 ;   n_3 \pm 1 , n_3 \pm \Delta x_3 \right)
\\ \nonumber
 \times T^{\prime \prime}(n_1 \pm \Delta x_1, n_1; n_2 \pm 
\Delta x_2, n_2; n_3 \pm \Delta x_3, n_3 )
\end{eqnarray}
In the above expression, the $T^{\prime \prime}$ term comes from the Graphene
Hamiltonian itself and the $T^{\prime}$ term comes due to putting the extra
term involving $H_T$ into the Hamiltonian.}.  
%%%%%%%%%%%%%%%%%%%%%%%%%%%%%%%%%%%%%%%%%%%%%%%%%%%%%%%%%%%%%%%%
\subsection{Plots for different coins}
Now for $S$, we can choose different types of mixing operators. 

\paragraph{Case 1 :} Let's take $S$ to be the three dimensional DFT coin. Therefore,
\begin{eqnarray}
S = \frac{1}{\sqrt{3}} \begin{pmatrix}
1 & 1 & 1 \\
1 & \omega & \omega^2 \\
1 & \omega^2 & \omega^4
\end{pmatrix}, \,\,\,\,\,\,\,\, \omega = \frac{-1+\sqrt{3}i}{2} \label{eq:dft_3d}
\end{eqnarray}
\begin{figure}[hbtp]
\begin{minipage}[c]{.5\linewidth}	
\centering
\includegraphics[scale=.26]{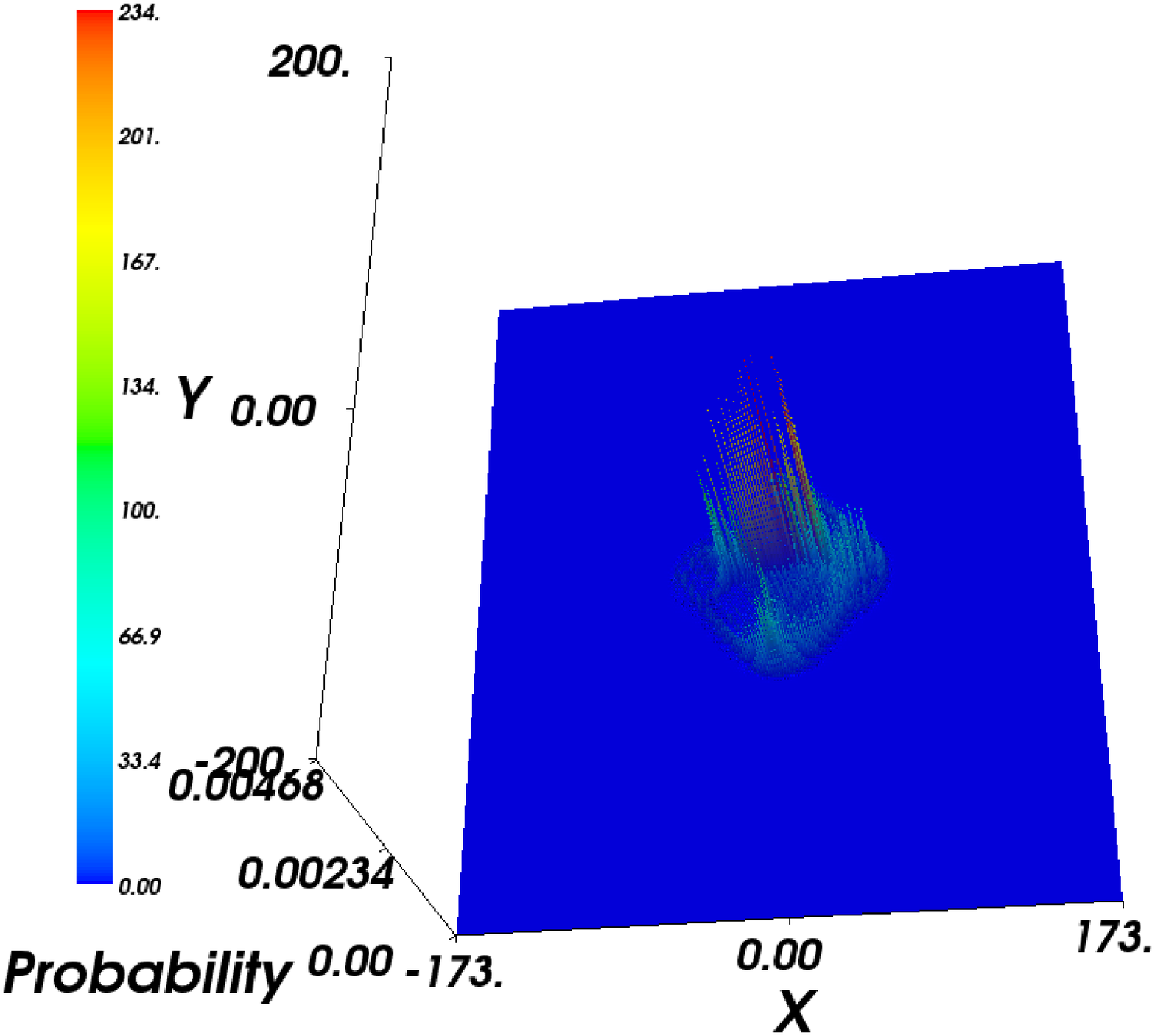}
\caption{Plot of the probability of finding the particle after 200 time steps using the single step approach and three dimensional DFT coin of Eq.\eqref{eq:dft_3d}. The values in the color-bar are multiplied by 50000. }
\label{graphene_dft_single_step}
\end{minipage}
\,\,\,\,\,\,
\begin{minipage}[c]{.5\linewidth}
\centering
\includegraphics[scale=.24]{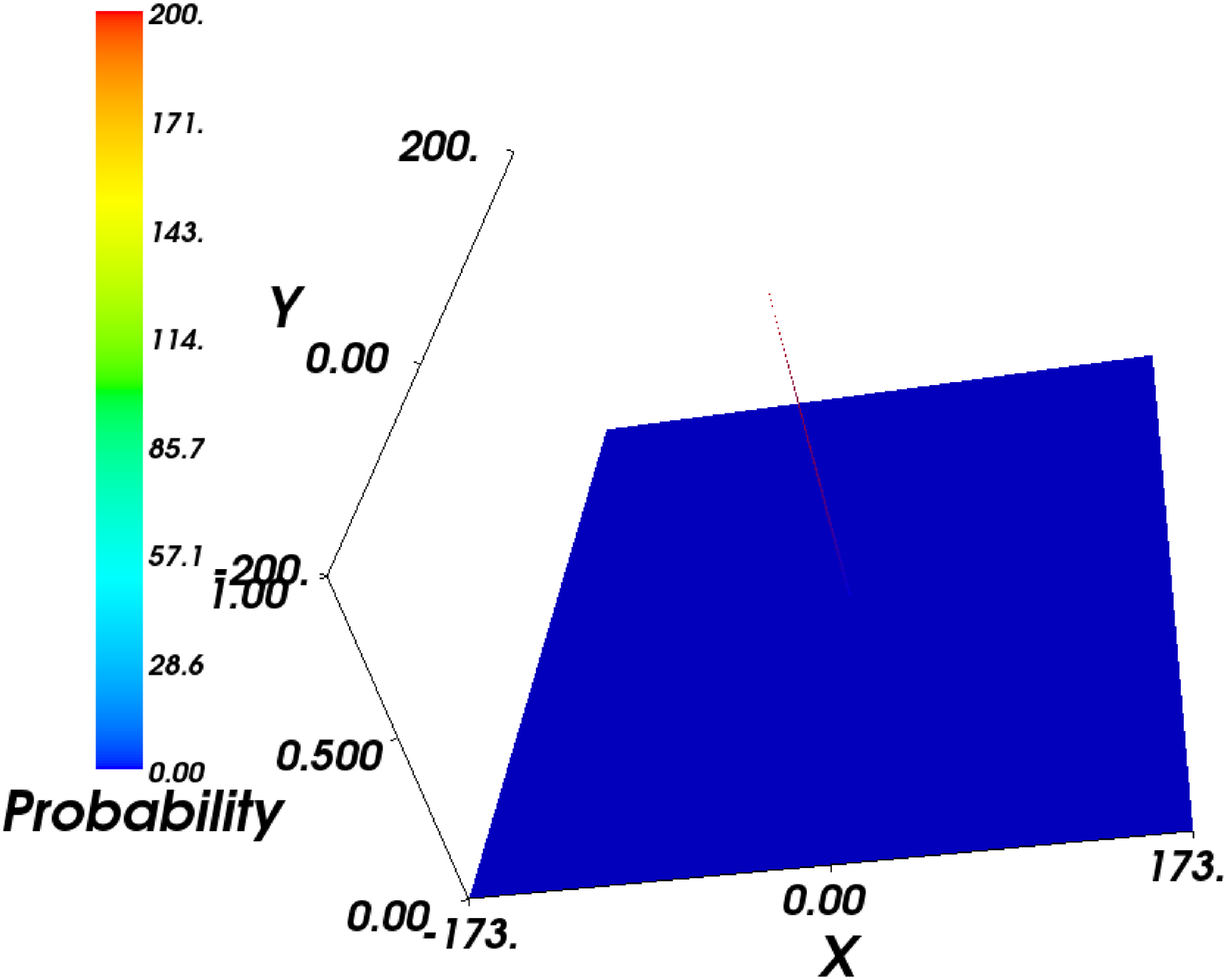}
\vspace{.2 cm} \caption{Plot of the probability of finding the particle after 200 time steps using the three step approach and the three dimensional DFT coin of Eq.\eqref{eq:dft_3d}. The values in the color-bar are multiplied by 200.}
\label{graphene_dft_three_step}
\end{minipage}
\end{figure}

We choose our initial chiral state to be $ \vert \chi \rangle = \frac{1}{\sqrt{3}}\left( \vert 1 \rangle + \vert 2 \rangle + \vert 3 \rangle \right) \otimes \vert - \rangle $. The results are shown in the following figures. The plot of Fig. \ref{graphene_dft_single_step} is generated by taking the additive time evolution operator (i. e. single step approach) and for that of \ref{graphene_dft_three_step}, we take the time evolution operator in a multiplicative form (which is equivalent to a three step approach here). 

In case of three step approach, we see that the probability of finding the particle collapses to a single point only. This is expected from the formulation of QRW, since for the multiplicative form of the evolution operator a particle starting from origin always reaches the point (1,-1,1) for every odd step and returns to the origin for every even step. 

\vspace{1 cm}

\paragraph{Case 2 :} Now let's choose $S$ to be the three dimensional Grover coin. Therefore,

\begin{figure}[hbtp]
\centering
\includegraphics[scale=.28]{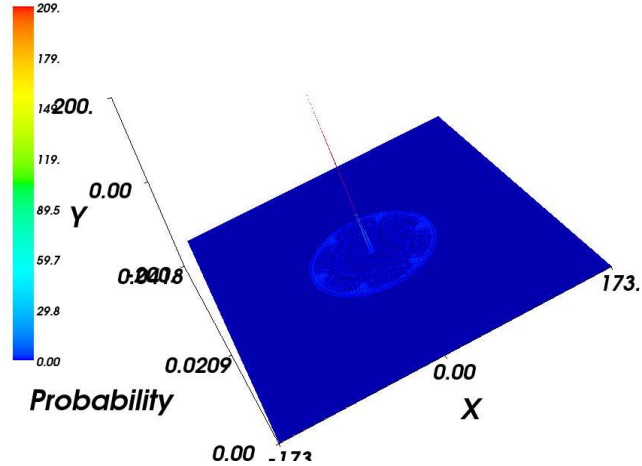}
\caption{Plot of the probability of finding the particle after 200 time steps using the single step approach and three dimensional Grover coin of Eq.\eqref{eq:grover_3d}. The values in the color-bar are multiplied by 5000.} \label{graphene_grover_single_step}
\end{figure}

\begin{eqnarray}
S = \frac{1}{3}
\begin{pmatrix}
-1 & 2 & 2 \\
2 & -1 & 2 \\
2 & 2 & -1
\end{pmatrix} \,\, . \label{eq:grover_3d}
\end{eqnarray}

The result for single step approach is shown in Fig. \ref{graphene_grover_single_step}. Here we have chosen our initial chiral state to be $ \vert \chi \rangle = \frac{1}{\sqrt{3}}\left( \vert 1 \rangle + i \vert 2 \rangle - i \vert 3 \rangle \right) \otimes \vert - \rangle $.  The plot for the three step approach is same as Fig. \ref{graphene_dft_three_step}. Therefore, we omit it for the sake of brevity.
%%%%%%%%%%%%%

\subsection{Calculation of variance}
\begin{figure}[hbtp]
\centering
\includegraphics[scale=.45]{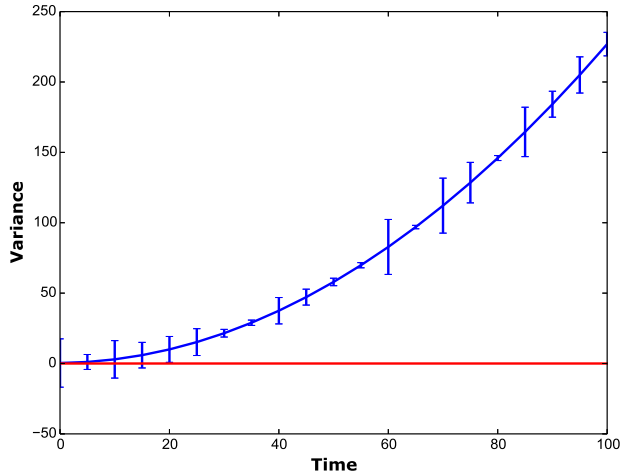}
\caption{Plot of variance with time for the quantum walk on Graphene
  lattice. The blue line shows the variance for the single step
  approach and the red one for the three step approach. In case of the
  single step approach, the errorbar w.r.t. the quadratic polynomial
  is multiplied by 50. Three dimensional DFT coin of
  Eq.\eqref{eq:dft_3d} has been used. } 
\label{variance_graphene}
\end{figure}

We calculated the variance for different time steps for both single step and three step approach using the coins of Eq.\eqref{eq:dft_3d} and Eq.\eqref{eq:grover_3d}. Since the results were similar, we include here only the result obtained with the three dimensional DFT coin. The plot with blue line in Fig.~\ref{variance_graphene} is for evolution operator in the additive form and that of the red line in the same figure corresponds to the time evolution operator in multiplicative form. For the single step approach, the relation of variance with time has an approximated polynomial form $ 0.34339904 + 0.03773854 t + 0.02227997 t^2 $. But for the three step approach, we see that the variance is zero, which is expected since the particle always reaches a single point in this case as mentioned previously.

%%%%%%%%%%%%%%%%%%%%%%%%%%%%%%%%%%%%%%%%%%%%%%%%%%%%%%%%%%%%%%%
 \section{Discussion}

Our approach to DTQRW sets constraints to the form of the Hamiltonian of the system. We found that the QW Hamiltonian can be described by the Weyl Hamiltonian with a
Dirac comb type potential. In Ref.~\citep{Chandrashekar_2013},
though it is suggested that the Hamiltonian for QW in one, two and
three dimension resembles a two component Dirac like Hamiltonian, our
form of the effective Hamiltonian is distinctly different from the
form of the Hamiltonian prescribed in the above reference. The other
important difference with Ref.~\citep{Chandrashekar_2013} for
multidimensional lattices, is the way in which the evolution operator
is perceived. In the referred work these evolution operators are
multiplicative while in our approach the evolution operator is
additive for dimensions higher than one. We have shown that the
multiplicative evolution operator can be produced from our formulation
also in case of two step approach by setting $k_x = 0$ or $k_y=0$ at
alternative lattice points.

The Fourier space formulation of QW as studied in Ref.~\citep{Nayak_Vishwanath_2000} gives an
analytic asymptotic form of the wave function of the particle. Hence
this method results in a good understanding of the problem. In this
approach, the method to solve the problem reduces to diagonalization
followed by an inverse Fourier transformation. It works very well in
case of 1D and 2D square lattice problem, but in case of more
complicated lattice like Graphene lattice, this method becomes too
complicated. In this situation, position space formulation turns out
to be more useful. Position space formulation gives a formal way for
numerical simulation of the problem of DTQRW for any kind of
lattice.

In this article, we have calculated the variance of the probability
distribution and found its quadratic dependence with number of
steps. With a little modification of the standard Graphene
Hamiltonian, we obtained the quantum walk on a Graphene lattice. Since
the Graphene lattice is more constrained, we expect more centred
probability distribution which is in agreement with our numerical
computation of the walk as shown in Fig.~\ref{graphene_dft_single_step}
and Fig.~\ref{graphene_grover_single_step}. Due to the geometry of the
Graphene lattice, $\vert +\rangle$ and $\vert - \rangle$ state can not
exist for the same lattice point. So there is no interference between
those states. Still in this case we find the quadratic dependence of
variance with time Fig.~\ref{variance_graphene}. The possible reason for
this quadratic dependence may be the interference between $\vert 1
\rangle$, $\vert 2 \rangle$ and $\vert 3 \rangle$ states. Use of
multiplicative time evolution operator in this case leads to the
confinement of the particle at a single point and thus no spreading
occurs as seen in Fig.~\ref{graphene_dft_three_step}. Hence we conclude that, the
additive time evolution operator is preferable over the multiplicative
one.

In short, we obtained an effective Hamiltonian for QW from the
evolution algorithms already existing in the literature. The form of the Hamiltonian became important
when the analysis on Graphene lattice was presented, because a standard
Hamiltonian of similar kind (linear in $k$) was already existing there. Using the similarity of the Graphene
Hamiltonian and the effective Hamiltonians for QW a new algorithm of
QW on Graphene lattice was presented. In a nutshell, we have presented a new
and interesting way of attaining an effective Hamiltonian for QW
which reproduces most of the results of earlier work in the same direction
in one-dimension but differs from earlier results in higher dimensions.
 
\section{Acknowledgements} 
N. P.  acknowledges  the financial support from the Council  of  Scientific  and  Industrial  Research  (CSIR),  India as a SPM JRF. 
 
%----------------------------------------------------------------------------------------
%	BIBLIOGRAPHY
%----------------------------------------------------------------------------------------

\label{Bibliography}

\bibliographystyle{unsrtnat} % Use the "unsrtnat" BibTeX style for formatting the 
\bibliography{graphene_bibliography.bib} 
% The references (bibliography) information are stored in the file named "Bibliography.bib"
%----------------------------------------------------

\begin{appendices}
\section{Derivation of Hamiltonian from Evolution operators in 1D} \label{append:deriv:H1D}

\subsection{Derivation of $H_T$} \label{append:deriv:HT1D}
If the Hamiltonian corresponding to $T$ is defined as $H_T \otimes \mathbbm{1}$, then it can derived as follows. 
\begin{eqnarray}
e^{-i H_T \Delta t \otimes \mathbbm{1}} &=& \sum_x \begin{pmatrix}
\vert x - \Delta x\rangle \langle x \vert & 0 \\
0 & \vert x + \Delta x \rangle \langle x \vert
\end{pmatrix} \nonumber \\
&=& \sum_x \begin{pmatrix}
e^{i k \Delta x} & 0 \\
0 & e^{- i k \Delta x}
\end{pmatrix}\left( \mathbbm{1}_2 \otimes \vert x \rangle \langle x \vert \right) \nonumber \\
&=& \begin{pmatrix}
e^{i k \Delta x} & 0 \\
0 & e^{- i k \Delta x}
\end{pmatrix} \,\,\,\,\,\,\,\, \left( \because \sum_x \vert x\rangle \langle x\vert = \mathbbm{1} \right) \nonumber  \\
\Rightarrow H_T \Delta t \otimes \mathbbm{1} &=& i \ln \begin{pmatrix}
e^{i k \Delta x} & 0 \\
0 & e^{- i k \Delta x}
\end{pmatrix}  \nonumber \\
&=& -k \Delta x \begin{pmatrix}
1 & 0 \\ 0 & -1
\end{pmatrix} \nonumber \\
\Rightarrow H_T \otimes \mathbbm{1} &=& - k v \sigma_z \,\,\,\,\,\,\,\,\,\,\,\,\, \left( \because v = \frac{\Delta x}{\Delta t} \right) \,\, . \nonumber
\end{eqnarray}
Thus we see that the derived Hamiltonian $H_T \otimes \mathbbm{1}$ is Hermitian.

\subsection{Derivation of $H_S$} \label{append:deriv:HS1D}

If we denote the Hamiltonian governing the chirality flip as $\sum_m H_S \otimes \vert m \rangle \langle m \vert$, we can write
\begin{eqnarray}
e^{-i \sum_m H_S \Delta \tau \otimes \vert m \rangle \langle m \vert} &=& \sum_m S \otimes \vert m \rangle \langle m \vert \nonumber \\
\Rightarrow e^{-i \sum_m H_S \Delta \tau \otimes \vert m \rangle \langle m \vert}\left( \mathbbm{1}_2 \otimes \vert n \rangle \right) &=& \sum_m S \otimes \vert m \rangle \langle m \vert \nonumber \left( \mathbbm{1}_2 \otimes \vert n \rangle \right) \\
\Rightarrow e^{-i H_S \Delta \tau} \otimes \vert n \rangle &=& S \otimes \vert n \rangle \nonumber \\
\Rightarrow H_S &=& \frac{i}{\Delta \tau} \ln S \,\, . \nonumber
\end{eqnarray}
Now $H_S$ is Hermitian by construction, since every unitary operator on a Hilbert space can be written as $U=e^{i A}$ for some Hermitian A.

\subsection{Derivation of H(x)} \label{append:H_x}

The total Hamiltonian $H$ is given by,
\begin{eqnarray}
H &=& \sum_n H_S \otimes \vert n\rangle \langle n \vert + H_T \otimes \mathbbm{1} \nonumber
\end{eqnarray}
Therefore,
\begin{eqnarray}
H(x) &=& \left(\mathbbm{1}_2 \otimes \langle x \vert \right) H \left( \mathbbm{1}_2 \otimes \vert \alpha \rangle \right) \,\,\,\, \left( \vert \alpha \rangle \,\, \textrm{is some ket} \in \mathbbm{R}, \textrm{the position space.} \right) \nonumber \\
&=& \sum_n H_S \delta (x-n) \langle n \vert \alpha \rangle + H_T \langle x \vert \alpha \rangle \nonumber \\
&=& \left( \sum_n H_S \delta (x-n) + H_T \right) \langle x \vert \alpha \rangle \nonumber \\
\Rightarrow H(x) &=& \sum_n H_S \delta (x-n) + H_T \,\, . \nonumber
\end{eqnarray}
Here in the $2^{nd}$ step we have used a proper unit weighting factor. 

\section{Derivation of Evolution operator from Hamiltonian in 2D square lattice} \label{append:deriv:W2D}

We have,

\begin{eqnarray}
H(n,m) &=& H_T \otimes \mathbbm{1} + H_S\otimes \vert n,m \rangle \langle n,m \vert \nonumber
\end{eqnarray}
Now following the logic as described in Sec. \ref{1D generalized hamiltonian}, we can say that $\left[H_S,H_T\right]\Delta t = 0$. Now,
\begin{eqnarray}
H_S &=& \frac{i}{\Delta \tau}\ln \left(S_x \otimes M_1 + S_y\otimes M_2 \right) \,\, , \nonumber \\
\Rightarrow e^{-i H_S \Delta \tau \otimes \vert n,m \rangle \langle n,m \vert} \left( \mathbbm{1}_4 \otimes \vert n,m \rangle \langle n,m \vert \right) &=&  \left(S_x \otimes M_1 + S_y\otimes M_2 \right) \otimes \vert n,m \rangle \langle n,m \vert \,\, . \nonumber
\end{eqnarray}
Again,
\begin{eqnarray}
H_T &=& -\left( k_x v_x \begin{pmatrix}
1 & 0 \\ 0 & 0
\end{pmatrix} + k_y v_y \begin{pmatrix}
0 & 0 \\ 0 & 1
\end{pmatrix}  \right) \otimes \sigma_z \,\, . \nonumber 
\end{eqnarray}
Therefore,
\begin{eqnarray}
e^{-i H_T \Delta t \otimes \mathbbm{1}} \left( \mathbbm{1}_4 \otimes \vert x,y \rangle \langle x,y \vert \right) &=& \left( \begin{pmatrix}
1 & 0 \\ 0 & 0
\end{pmatrix} \otimes \begin{pmatrix}
e^{i k_x \Delta x} & 0 \\ 0 & e^{-i k_x \Delta x}
\end{pmatrix} + \begin{pmatrix}
0 & 0 \\ 0 & 1
\end{pmatrix} \otimes \begin{pmatrix}
e^{i k_y \Delta y} & 0 \\ 0 & e^{-i k_y \Delta y}
\end{pmatrix}  \right) \otimes \vert x,y \rangle \langle x,y \vert \nonumber \\
\Rightarrow T(x,y) &=& \begin{pmatrix}
e^{i k_x \Delta x} & 0 & 0 & 0 \\
0 & e^{- i k_x \Delta x} & 0 & 0 \\
0 & 0 & e^{i k_y \Delta y} & 0 \\
0& 0 & 0 & e^{- i k_y \Delta y}   
\end{pmatrix} \otimes \vert x,y \rangle \langle x,y \vert \,\, . \nonumber
\end{eqnarray}
Therefore, we have,
\begin{eqnarray}
W(n,m) &=& T(n,m) \left( (S_x \otimes M_1 + S_y \otimes M_2) \otimes \vert n,m \rangle \langle n,m \vert \right) \,\, . \nonumber
\end{eqnarray}
\end{appendices}

\end{document}